\pdfoutput=1
\documentclass[12pt]{article}
\usepackage[utf8]{inputenc}
\usepackage{amsmath, amssymb}
\usepackage{graphicx}
\usepackage{booktabs}
\usepackage{placeins}
\usepackage{subcaption}
\usepackage{bbm}
\usepackage{pgfplots}
\pgfplotsset{compat=1.17}
\usepackage{hyperref}
\usepackage{parskip}

\title{Scaling Laws for Economic Productivity: Experimental Evidence in LLM-Assisted Consulting, Data Analyst, and Management Tasks\thanks{I gratefully acknowledge financial support from Coefficient Giving to conduct this research. The research described in this article was approved by the Yale Human Research Protection Program and was preregistered at the AEA RCT Registry (AEARCTR-0013743). We thank Yejun Yun for his invaluable research assistance on this experiment and Pascual Restrepo, Jasan Abaluck, and Tom Cunningham for comments and suggestions.}}
\author{
    Ali Merali\textsuperscript{\textdagger} \\ 
}

\begin{document}
\maketitle

\renewcommand{\thefootnote}{\textdagger} 
\footnotetext{Yale University, Department of Economics: ali.merali@yale.edu}

\renewcommand{\thefootnote}{\arabic{footnote}} 

\begin{abstract}
This paper derives ``Scaling Laws for Economic Impacts''- empirical relationships between the training compute of Large Language Models (LLMs) and professional productivity. In a preregistered experiment, over 500 consultants, data analysts, and managers completed professional tasks using one of 13 LLMs. We find that each year of model progress reduced task time by 8\%, with 56\% of gains driven by increased compute and 44\% by algorithmic progress. However, productivity gains were significantly larger for non-agentic analytical tasks compared to agentic workflows requiring tool use. These findings suggest continued model scaling could boost U.S. productivity by approximately 20\% over the next decade.

\end{abstract}

\pagebreak

\section{Introduction} 

Between the release of GPT-2 in 2019 and the frontier models of 2025, the amount of compute used to train large language models (LLMs) increased by approximately four orders of magnitude. The machine learning literature has derived remarkably consistent "scaling laws" from this explosion in resources, demonstrating that model performance—measured by cross-entropy loss—improves as a predictable power law of compute, data, and parameter size (Kaplan et al., 2020). Yet, for economists and policymakers, the critical question remains unanswered: how does a reduction in a model’s mathematical loss translate into tangible economic productivity? While the technological frontier is advancing rapidly, we possess little rigorous evidence on the elasticity of human professional output with respect to these model capabilities. 

To address this, we conducted a large-scale randomized controlled trial (RCT) involving over 500 professionals across three high-skill domains: management, data analysis, and consulting. Participants were tasked with completing complex workflows designed to be representative of their professions—ranging from strategic report writing and statistical hypothesis testing to tasks requiring multi-step tool use such as creating presentation slides or Gantt charts. Workers were randomly assigned to a control group or to a treatment group equipped with one of thirteen different LLMs, spanning various compute scales and release dates. We utilized high-powered incentives, including bonus payments that doubled base earnings for high-quality submissions as evaluated by expert peer graders.

Our primary contribution is the derivation of "Scaling Laws for Economic Impacts"—quantifying the relationship between model inputs and professional productivity. We decompose the progress of frontier AI into two distinct factors: the scaling of training compute and algorithmic innovation (e.g., architectural improvements and better training data). First, we identify a robust "calendar-time" scaling effect—which captures the aggregate of both factors—where each year of frontier model progress is associated with an 8\% reduction in task completion time ($p < 0.05$). Second, isolating the effect of scale, we find that a tenfold (10x) increase in model training compute is associated with a 6.3\% reduction in time taken. When decomposing these gains, we find that approximately 44\% of the observed improvement is attributable to algorithmic progress over time, while the remainder is driven by pure compute scaling.

We further establish the baseline "AI premium" by pooling all treatment groups. Access to any AI model increased base Earnings Per Minute (EPM) by 81.3\% ($p = 0.001$) and raised expert-assessed quality by 0.34 standard deviations. The compounding effects of speed and quality resulted in an 146\% increase in Total Earnings Per Minute (TEPM), inclusive of performance bonuses with almost equal contributions for this increase coming from greater speed (52.6\%) and higher quality (47.4\%).

However, we uncover significant heterogeneities in these gains across task types. While AI assistance delivered total earnings gains of \$1.58 per minute on non-agentic, analytical tasks, the gain fell to just \$0.34 per minute for "agentic" tasks requiring multi-step interactions with external tools—a disparity significant at the 5\% level ($p = 0.043$). This suggests that while current scaling paradigms are rapidly commoditizing analytical cognition, the productivity frontier for tasks requiring procedural agency currently remains significantly more resistant to automation \footnote{An important caveat is that participants were restricted to standard chatbot interfaces with text-based input and output and limited tool access. Consequently, this design may underestimate the capabilities of current AI models, particularly regarding agentic tasks.}.

Next, we investigate scaling laws for quality. We find a striking divergence: while the quality of autonomous model output scales linearly with compute, the quality of human-assisted output remains stagnant across model generations. This implies that human users effectively cap the realized capabilities of frontier models, satisfying to a fixed quality threshold rather than maximizing the tool's potential.

Finally, we utilize these experimental elasticities within an aggregate growth framework (Acemoglu, 2024). We estimate that continued model scaling could boost U.S. productivity by approximately 20\% over the next decade, assuming marginal costs of inference remain low. This figure significantly exceeds prior conservative estimates by explicitly incorporating the dynamic gains from predictable advancements in model compute, rather than treating AI capabilities as fixed.

The remainder of the paper is organized as follows. Section 2 reviews the related literature and Section 3 then outlines the experimental methodology. Section 4 presents the experimental results, establishing the baseline AI premium, deriving the economic scaling laws, and decomposing the drivers of productivity growth. Section 5 utilizes these elasticities to estimate aggregate productivity gains for the U.S. economy and Section 6 then concludes. Regression outputs for all figures shown as well as supplementary regressions are available in Appendix A and all the tasks completed by experiment participants can be found in Appendix B.

\section{Related Literature}

This paper contributes to three emerging strands of literature at the intersection of artificial intelligence and labor economics. First, we build upon a rapidly expanding body of experimental evidence documenting the productivity ``uplift'' of generative AI across diverse domains. Significant productivity gains have been documented for legal analysis by Choi, Monahan, and Schwarcz (2024), and for software engineering by Peng et al. (2023) and Cui et al. (2025). In the domain of high-skill knowledge work, Dell’Acqua et al. (2023) find substantial quality and speed improvements for management consultants. Similar effects are observed for professional writing by Noy and Zhang (2023), call center operations by Brynjolfsson, Li, and Raymond (2025), translation by Merali (2024), and entrepreneurship by Otis et al. (2024). Our work unifies these domain-specific findings by applying a consistent experimental design across multiple professions and, crucially, across thirteen distinct models to measure the elasticity of these uplifts with respect to model capabilities.

Second, we connect the economic literature to the study of AI scaling and evaluation. While Kaplan et al. (2020) and Hoffmann et al. (2022) have established rigorous scaling laws for model perplexity, the relationship between model size and downstream economic utility has remained under-explored. Our approach responds to calls from the machine learning community for more rigorous ``Centaur Evaluations''---benchmarks where humans and AI solve tasks cooperatively---as advocated by Haupt and Brynjolfsson (2025). By measuring the joint productivity of the human--AI system, we provide an economic counterweight to standard ``imitation games'' that test models in isolation. Furthermore, our findings speak to critiques of benchmark-centric evaluation by Kulveit et al. (2025) and Raji et al. (2021), who argue that leaderboard-style, model-in-isolation benchmarks can misrepresent real-world performance and can steer research away from human--AI complementarity in realistic workflows.

Finally, this paper contributes to the development of economically grounded benchmarks for AI progress. A new wave of evaluations has attempted to quantify the potential economic utility of frontier models, including OpenAI's ``GDPval'' (OpenAI, 2025), Mercor's ``AI Productivity Index'' (APEX) (Vidgen et al., 2025), and technical benchmarks such as SWE-bench (Jimenez et al., 2024) and GAIA (Mialon et al., 2023). However, these frameworks generally evaluate model outputs in isolation (even when graded by experts or judge models), rather than measuring the productivity of human--AI teams. By contrast, our results constitute a benchmark of human-augmented productivity. By quantifying the elasticity of real-world economic output---measured via both speed and quality---with respect to model capabilities, we provide a direct metric of the realized marginal productivity of labor in the presence of frontier technology. This offers a necessary complement to automated evaluations, ensuring that economic forecasting is grounded in the actual, rather than theoretical, deployment of frontier models.

\section{Experimental Design}

For this experiment, over 500 participants were recruited, primarily through the online research platform Prolific, with a smaller subset of experienced professionals recruited directly. These participants were stratified roughly evenly across three high-skill domains: Management, Consulting, and Data Analysis. All participants underwent a rigorous multi-stage screening process to ensure professional competency. Eligibility requirements included a self-reported annual salary exceeding \$40,000, at least one year of professional experience in their respective field, and high self-rated proficiency in key professional skills. Additionally, participants were required to pass a demanding screening survey containing strict attention checks and competency verifications, which resulted in a screen-out rate of approximately 90\% of initial applicants.

The final pool of participants was highly experienced. Analysis of the sample indicates that over 79\% of participants had at least three years of professional experience, with nearly 47\% possessing five or more years of experience in their field. In self-reported assessments, participants rated their overall professional abilities at an average of 4.13 out of 5. Furthermore, they reported a moderate familiarity with AI tools, rating their familiarity at 3.62 out of 5 and their ability to use such tools at 3.56 out of 5.

Participants completed one of three tasks specific to their profession during the experiment. These tasks were designed to simulate real-world workflows and were categorized into ``Agentic'' tasks (requiring multi-step reasoning and tool use) and ``Non-Agentic'' tasks (focused on analysis and writing). The tasks covered a diverse range of professional activities, ranging from revising financial expansion reports and creating presentation slides to conducting statistical A/B test analyses and evaluating vendor contracts. A full description of the tasks can be found in Appendix B. On average, participants took approximately 26 minutes to complete each task and rated the difficulty of these tasks at 3.18 out of 5.

Participants were given high-powered incentives to ensure high-quality effort. The base payment for completing a task was set at \$15. However, participants could earn a substantial bonus of an additional \$15 (doubling their total pay to \$30) if their submission received a grade of 5, 6, or 7 out of 7. Grades were assigned by expert peer graders with over five years of professional experience in the relevant field. Submissions receiving a score of 0 or 1 out of 7---typically indicating a failure to follow instructions or evidence of bot usage---were not compensated; such instances of non-compliance occurred in less than 5\% of the sample.

To assist with their tasks, participants were randomly assigned to a treatment group with access to one of thirteen specific AI models, or to a control group. Access was provided through a custom-built website that allowed for the monitoring of model usage. Prior to the main tasks, participants completed a monitored practice attempt to familiarize themselves with the assigned AI model and ensure compliance with experimental protocols. Participants in the treatment group reported high engagement with the tools, rating their usage of the bot at an average of 3.59 out of 5 and its helpfulness at 3.58 out of 5.

\section{Results}

\textbf{Section 4.1: Productivity and Quality Impacts of any AI Usage} 

We begin by establishing the baseline impact of AI assistance on worker productivity and output quality across all professions in our sample (Consultants, Data Analysts, and Managers). To do so, we pool the results from all thirteen AI models against the control group. The data reveals substantial productivity gains across four key dimensions. 

We find that AI usage drives a statistically significant increase in base Earnings Per Minute (EPM) of \$0.56 ($p=0.001$), representing a 81.3\% increase relative to the control group baseline. Simultaneously, AI assistance raised output quality by 0.6 points ($p <0.001$), an increase of 18\% or 0.34 standard deviations. When accounting for the performance bonuses awarded for high-scoring tasks (bonus payments doubled earnings and were awarded for grades of 5 or higher out of seven), this dual improvement in speed and quality compounded to raise Total Earnings Per Minute (TEPM) by \$1.06 ($p <0.001$) - a 146\% increase in overall economic value. Finally, we conclude by demonstrating that these aggregate premiums are driven by specific task types, distinguishing between the impacts on agentic versus non-agentic workflows.

First, we show the effect of AI on Earnings per Minute through faster task completions. For tasks meeting the minimum quality threshold (a grade of at least 2 out of 7), the average earnings rose from \$0.69 per minute for the control group to \$1.24 per minute for the treatment group as depicted by Figure 1 below ($p <0.001$) with full regression results available in Table 1 in Appendix A (all results on Time Taken are also in Appendix A). 

\begin{figure}[htbp]
    \centering
    \textbf{Figure 1: Impact on Earnings Per Minute of AI Usage}\\[0.75em]
    \label{fig:anyai_epm_bar}
    \begin{tikzpicture}
    \begin{axis}[
        ybar,
        bar width=0.55cm,
        symbolic x coords={Without AI, With AI},
        xtick=data,
        ymin=0,
        ylabel={Earnings per minute (\$/min)},
        enlarge x limits=0.25,
        width=0.6\textwidth,
        height=7cm,
        ymajorgrids=true,
        grid style={dashed,gray!40},
        ticklabel style={font=\small},
        label style={font=\small},
        nodes near coords,
        every node near coord/.append style={font=\small},
        nodes near coords align={vertical},
        /pgf/number format/fixed,
        /pgf/number format/precision=2
    ]
        \addplot coordinates {(Without AI,0.69) (With AI,1.24)};
    \end{axis}
    \end{tikzpicture}
\end{figure}

Secondly, we examine how AI assistance affects output quality. Excluding only tasks that received a grade of zero, we find that access to any AI model substantially raises expert-assessed grades. In regression-adjusted terms (controlling for profession and task), the average grade increases from 3.32 without AI to 3.91 with AI, an improvement of 0.60 points on a seven-point scale ($p<0.001$), corresponding to roughly one-third of a standard deviation. These results, summarized in Figure 2, indicate that the productivity gains documented above do not come at the expense of lower quality; on the contrary, AI assistance simultaneously accelerates task completion and improves output quality. Full regression results are reported in Table 2 in Appendix A.

\begin{figure}[htbp]
    \centering
    \textbf{Figure 2: Impact on Grades of AI Usage}\\[0.75em]
    \begin{tikzpicture}
    \begin{axis}[
        ybar,
        bar width=0.55cm,
        symbolic x coords={Without AI, With AI},
        xtick=data,
        ymin=0,
        ymax=7,
        ylabel={Mean grade (0--7)},
        enlarge x limits=0.25,
        width=0.6\textwidth,
        height=7cm,
        ymajorgrids=true,
        grid style={dashed,gray!40},
        ticklabel style={font=\small},
        label style={font=\small},
        nodes near coords,
        every node near coord/.append style={font=\small},
        nodes near coords align={vertical},
        /pgf/number format/fixed,
        /pgf/number format/precision=2
    ]
        \addplot coordinates {(Without AI,3.32) (With AI,3.91)};
    \end{axis}
    \end{tikzpicture}
    \label{fig:anyai_grade_bar}
\end{figure}

Thirdly, we consider the impact of AI assistance on Total Earnings Per Minute (TEPM), which incorporates both base pay and performance bonuses. Workers earned a fixed \$15 per task conditional on meeting the minimum quality threshold (grade $\geq 2$), and an additional \$15 bonus (i.e.\ earnings doubled) for high-quality submissions with grades of 5 or higher. Using the same paid-task sample and controlling for profession and task, AI assistance raises TEPM from \$0.73 per minute to \$1.79 per minute, an increase of \$1.06 ($p<0.001$), corresponding to an 146\% increase in total earnings per unit of time. Figure 3 summarizes these regression-adjusted means, with full regression results reported in Table 3 in Appendix A.

\begin{figure}[htbp]
    \centering
    \textbf{Figure 3: Impact on Total Earnings Per Minute (TEPM) of AI Usage}\\[0.75em]
    \begin{tikzpicture}
    \begin{axis}[
        ybar,
        bar width=0.55cm,
        symbolic x coords={Without AI, With AI},
        xtick=data,
        ymin=0,
        ylabel={Total earnings per minute (\$/min)},
        enlarge x limits=0.25,
        width=0.6\textwidth,
        height=7cm,
        ymajorgrids=true,
        grid style={dashed,gray!40},
        ticklabel style={font=\small},
        label style={font=\small},
        nodes near coords,
        every node near coord/.append style={font=\small},
        nodes near coords align={vertical},
        /pgf/number format/fixed,
        /pgf/number format/precision=2
    ]
        \addplot coordinates {(Without AI,0.73) (With AI,1.79)};
    \end{axis}
    \end{tikzpicture}
    \label{fig:anyai_tepm_bar}
\end{figure}

Taken together, these estimates imply that the overall TEPM premium of \$1.06 per minute from AI assistance reflects both faster task completion and higher bonus rates. Roughly \$0.56 per minute (52.6\% of the total) comes from higher base earnings per minute (EPM), while the remaining \$0.50 per minute (47.4\%) is driven by improved quality that pushes more tasks above the bonus threshold. AI thus delivers economically meaningful gains along both the speed and quality margins, rather than merely shifting the composition of pay.

Finally, we examine how these effects vary across task types. We classify tasks as agentic when successful completion requires multi-step external actions and tool use (e.g.\ creating Gantt charts in spreadsheet software and sending them via email, or extracting information from multiple 100+ page PDFs and running statistical tests). In contrast, non-agentic tasks are primarily analytical or interpretive (e.g.\ hypothesis testing, interpreting results, or writing analytical reports).

Using the same specifications as above, we find much larger gains from AI usage on non-agentic tasks (regression outputs shown in Table 4 and Table 5 of Appendix A). For Earnings Per Minute, the effect of any AI usage is \$0.83 per minute on non-agentic tasks ($p<0.001$) but only \$0.18 per minute on agentic tasks ($p=0.48$), with the difference between the two significant at the 5\% level ($p=0.050$). For grades, AI raises quality by 0.82 points on a seven-point scale on non-agentic tasks ($p<0.001$), compared to 0.27 points on agentic tasks ($p=0.31$); the difference is not statistically significant ($p=0.11$). 

Finally, Total Earnings Per Minute displays the sharpest heterogeneity: AI increases total earnings by \$1.58 per minute on non-agentic tasks ($p<0.001$), but only \$0.34 per minute on agentic tasks ($p=0.46$), with the difference between these effects significant at the 5\% level ($p=0.043$). Figure 4 illustrates this heterogeneity in total earnings, and Table 6 in Appendix A reports the corresponding regressions.

\begin{figure}[htbp]
    \centering
    \textbf{Figure 4: Heterogeneity in TEPM Gains by Task Type}\\[0.75em]
    \begin{tikzpicture}
    \begin{axis}[
        ybar,
        bar width=0.55cm,
        symbolic x coords={Non-agentic, Agentic},
        xtick=data,
        ymin=0,
        ylabel={AI effect on TEPM (\$/min)},
        enlarge x limits=0.25,
        width=0.6\textwidth,
        height=7cm,
        ymajorgrids=true,
        grid style={dashed,gray!40},
        ticklabel style={font=\small},
        label style={font=\small},
        nodes near coords,
        every node near coord/.append style={font=\small},
        nodes near coords align={vertical},
        /pgf/number format/fixed,
        /pgf/number format/precision=2
    ]
        \addplot coordinates {(Non-agentic,1.58) (Agentic,0.34)};
    \end{axis}
    \end{tikzpicture}
    \label{fig:agentic_tepm_bar}
\end{figure}

\textbf{Section 4.2: Scaling Laws for Economic Productivity} 

The previous section compared workers who received access to any AI model with those in the control group and documented large average gains in productivity and quality. In this section, we move beyond this binary comparison and start to unpack what drives variation in these gains across models. Rather than treating all AI assistance as homogeneous, we ask how economic outcomes change as the underlying capabilities of the assigned model improve. 

Conceptually, two inputs underlie frontier LLM progress. One is sheer scale: increases in training compute, which the machine-learning literature has shown to be tightly linked to model performance. The other is algorithmic progress: improvements in architectures, training procedures, data, and alignment that arrive over calendar time even holding compute fixed. We proxy this second component using the model’s release date, measured as months since November 2022 (the public launch of ChatGPT), assigning zero months to the no-bot control. 

Our empirical strategy therefore traces out “economic scaling laws” along two related axes. First, we estimate how log time taken, earnings per minute (EPM), and total earnings per minute (TEPM) vary with model release month, interpreting the slope as the average change in economic outcomes per year of frontier progress. We then allow these calendar-time slopes to differ by profession and by task type (agentic versus non-agentic). Second, we replace calendar time with log training compute and recover elasticities with respect to a tenfold increase in compute. Finally, we combine both months and log compute in the same specification to shed light on how much of the observed improvement in economic outcomes is associated with algorithmic progress over time versus simple scaling of compute. 

We first ask whether the economic gains from AI assistance improve as frontier models become more capable over calendar time. The magnitudes are economically large. Moving one year forward in model release date is associated with roughly an 8\% reduction in time taken to complete a task as shown in Figure 5 below ($p \approx 0.04$, full regression results and a breakdown of the scaling law by profession can be found in Table 9 and Figure A2 of Appendix A respectively)\footnote{The annual effect is computed as $100 \times (\exp(12 \hat{\beta}) - 1)$, where $\hat{\beta}$ is the monthly coefficient on months since November 2022.}. Adding the full set of background controls (demographics, abilities, AI familiarity and use, and country dummies) leaves this estimate essentially unchanged. Given that the average task in the human-only group takes around 25 minutes, this corresponds to workers completing the same tasks several minutes faster per year purely from frontier progress in the underlying models.

\begin{figure}[htbp]
    \centering
    \textbf{Figure 5: Scaling Laws for Time Taken over Calendar Time}\\[0.75em]
    \begin{tikzpicture}
    \begin{axis}[
        xlabel={Model release date},
        ylabel={Log time taken},
        xmin=0, xmax=36,
        ymin=6, ymax=7.6,
        xmajorgrids=true,
        ymajorgrids=true,
        grid style=dashed,
        height=9cm,
        width=0.8\textwidth,
        xtick={0,12,24,36},
        xticklabels={Nov 22, Nov 23, Nov 24, Nov 25},
        ticklabel style={font=\small},
        label style={font=\small},
        legend style={font=\small, at={(0.98,0.02)},anchor=south east},
        legend cell align={left}
    ]

        \addplot[
            only marks,
            mark=x,
            mark size=3pt,
            color=blue,
            forget plot 
        ] coordinates {
            (0.00,7.091)  
            (2.99,6.930)  
            (7.56,7.056)  
            (9.89,7.104)  
            (14.52,6.897) 
            (15.41,6.808) 
            (16.59,6.939) 
            (16.59,7.249) 
            (17.41,6.879) 
            (18.66,6.825) 
            (21.42,6.699) 
            (26.94,7.189) 
            (27.76,6.875) 
            (27.79,7.295) 
            (28.52,6.942) 
            (32.23,6.856) 
        };

        \addplot[
            domain=0:36,
            samples=100,
            color=red,
            forget plot
        ] {7.045 - 0.0071 * x};

        \addplot[
            domain=0:36,
            samples=2,
            dashed,
            color=blue
        ] {7.091};
        \addlegendentry{Mean Time for Human-Only Participants}

        \node at (axis cs: 12,7.4) [anchor=west, text=red]
            {$\approx 8\%$ reduction in time per year};

    \end{axis}
    \end{tikzpicture}
    \label{fig:logtime_months}
\end{figure}

Given the substantial reductions in task completion time documented above, it is perhaps unsurprising that we observe a corresponding scaling law for Earnings Per Minute (EPM). As base pay is fixed per task (conditional on meeting the quality threshold) and excludes performance bonuses, EPM is functionally the inverse of time taken. As illustrated in Figure 6, the economic value generated by workers scales linearly with the release date of the model they utilize. Our regression estimates indicate that each month of frontier model progress is associated with an increase in base earnings of \$0.019 per minute ($p < 0.01$). Aggregated over a year, this implies that the algorithmic progress and increased model training compute raises the value of professional output by approximately \$14 per hour annually. These results, detailed in Table 10 in Appendix A, confirm that the time-savings afforded by newer models translate directly into higher hourly productivity rates.

\begin{figure}[htbp]
    \centering
    \textbf{Figure 6: Scaling Laws for Earnings over Calendar Time}\\[0.75em]
    \begin{tikzpicture}
    \begin{axis}[
        xlabel={Model release date},
        ylabel={Earnings per minute (\$/min)},
        xmin=0, xmax=36,
        ymin=0.6, ymax=2.6,
        xmajorgrids=true,
        ymajorgrids=true,
        grid style=dashed,
        height=9cm,
        width=0.8\textwidth,
        xtick={0,12,24,36},
        xticklabels={Nov 22, Nov 23, Nov 24, Nov 25},
        ticklabel style={font=\small},
        label style={font=\small},
        legend style={font=\small, at={(0.98,0.02)},anchor=south east},
        legend cell align={left}
    ]

        \addplot[
            only marks,
            mark=x,
            mark size=3pt,
            color=blue,
            forget plot
        ] coordinates {
            (0.00, 0.995)
            (2.99, 1.360)
            (7.56, 1.249)
            (9.89, 1.111)
            (14.52, 1.767)
            (15.11, 1.816)
            (16.59, 1.580)
            (16.59, 1.092)
            (17.41, 1.179)
            (21.42, 2.318)
            (26.84, 1.375)
            (26.94, 0.876)
            (27.76, 1.666)
            (27.80, 1.007)
            (28.52, 1.163)
            (32.23, 1.556)
        };

        \addplot[
            domain=0:36,
            samples=100,
            color=red,
            forget plot
        ] {0.9447 + 0.0194 * x};

        \addplot[
            domain=0:36,
            samples=2,
            dashed,
            color=blue
        ] {0.995};
        \addlegendentry{Mean EPM for Human-Only Participants}

        \node at (axis cs: 4,2.45) [anchor=west, text=red]
            {$\approx \$13.97$/hour increase in earnings per year};

    \end{axis}
    \end{tikzpicture}
    \label{fig:epm_months}
\end{figure}

Next, we turn to the most comprehensive metric of economic productivity: Total Earnings Per Minute (TEPM). As shown in Figure 7, the economic scaling law is steepest here: our estimates indicate that for every month of frontier model progress, a worker’s total earnings capacity increases by \$0.037 per minute ($p < 0.001$), or an effective hourly wage increase of approximately \$26.30 per year. However, it is crucial to interpret the magnitude of this slope in the context of the incentive structure, where the performance bonus (for submissions with grades 5 or higher out of 7) doubles the effective piece rate. We observe that the TEPM coefficient (0.037) is approximately twice the magnitude of the base EPM coefficient (0.019), both figures reported in Table 10 and 11 of Appendix A respectively. This suggests that the compounding value of frontier models is driven primarily by the acceleration of workflows—simply allowing workers to complete high-value, bonus-eligible tasks at a faster rate—rather than by a distinct scaling law that drastically alters the probability of achieving a bonus.

\begin{figure}[htbp]
    \centering
    \textbf{Figure 7: Scaling Laws for Total Earnings Per Minute over Calendar Time}\\[0.75em]
    \begin{tikzpicture}
    \begin{axis}[
        xlabel={Model release date},
        ylabel={Total earnings per minute (\$/min)},
        xmin=0, xmax=36,
        ymin=1.0, ymax=4.2,
        xmajorgrids=true,
        ymajorgrids=true,
        grid style=dashed,
        height=9cm,
        width=0.8\textwidth,
        xtick={0,12,24,36},
        xticklabels={Nov 22, Nov 23, Nov 24, Nov 25},
        ticklabel style={font=\small},
        label style={font=\small},
        legend style={font=\small, at={(0.98,0.02)},anchor=south east},
        legend cell align={left}
    ]

        \addplot[
            only marks,
            mark=x,
            mark size=3pt,
            color=blue,
            forget plot
        ] coordinates {
            (0.00, 1.319)
            (2.99, 1.500)
            (7.56, 1.850)
            (9.89, 1.605)
            (14.52, 3.142)
            (15.11, 3.632)
            (16.59, 2.160)
            (16.59, 1.658)
            (17.41, 1.808)
            (21.42, 3.777)
            (26.84, 2.222)
            (26.94, 1.753)
            (27.76, 2.678)
            (27.80, 1.616)
            (28.52, 1.651)
            (32.23, 1.743)
        };

        \addplot[
            domain=0:36,
            samples=100,
            color=red,
            forget plot
        ] {1.0292 + 0.0365 * x};

        \addplot[
            domain=0:36,
            samples=2,
            dashed,
            color=blue
        ] {1.319};
        \addlegendentry{Mean TEPM for Human-Only Participants}

        \node at (axis cs: 6.5,3.95) [anchor=west, text=red]
            {$\approx \$26.30$/hour increase per year };

    \end{axis}
    \end{tikzpicture}
    \label{fig:tepm_months}
\end{figure}

Finally, disaggregating these results by task structure reveals two distinct scaling laws. As shown in Figure 8, both task types benefit from frontier model progress, but the rate of improvement differs significantly. For non-agentic tasks, the scaling curve is steep: the estimated coefficient implies that task completion time falls by approximately 10.7\% per year of frontier model progress ($p < 0.05$). For agentic tasks, while the sign remains positive, the slope is considerably flatter (approximately 4.8\% reduction per year) and statistically indistinguishable from zero in this sample. Similar results are shown for earnings per minute in Figure 9, below. This divergence suggests that while scaling effectively compresses the time required for well-defined analytical and managerial workflows, the complex planning and multi-step reasoning required for agentic tasks may face more stubborn bottlenecks that recent model improvements have not yet fully overcome. Full regression results for this split are provided in Tables \ref{tab:logtime_months_nonagentic} and \ref{tab:logtime_months_agentic} of Appendix A.

\begin{figure}[htbp]
    \centering
    \textbf{Figure 8: Scaling Laws for Time Taken over Calendar Time, by Task Type}\\[0.75em]

\begin{subfigure}[t]{0.49\textwidth}
    \centering
    \textbf{Non-agentic tasks}\\[0.25em]
    \begin{tikzpicture}
    \begin{axis}[
        xlabel={Model release date},
        ylabel={Log time taken},
        xmin=0, xmax=36,
        ymin=6, ymax=8.1,
        xmajorgrids=true,
        ymajorgrids=true,
        grid style=dashed,
        height=7.8cm,
        width=\textwidth,
        xtick={0,12,24,36},
        xticklabels={Nov 22, Nov 23, Nov 24, Nov 25},
        ticklabel style={font=\small},
        label style={font=\small},
        legend style={font=\tiny, at={(0.98,0.02)},anchor=south east},
        legend cell align={left}
    ]

        \addplot[
            only marks,
            mark=x,
            mark size=3pt,
            color=blue,
            forget plot
        ] coordinates {
            (0.00, 6.917) (2.99, 6.463) (7.56, 6.782) (9.89, 7.066)
            (14.52, 6.433) (15.11, 6.354) (16.59, 6.544) (16.59, 6.722)
            (17.41, 6.622) (21.42, 6.241) (26.84, 6.629) (26.94, 6.444)
            (27.76, 6.689) (27.80, 7.518) (28.52, 6.777) (32.23, 6.540)
        };

        \addplot[
            domain=0:36,
            samples=100,
            color=red,
            forget plot
        ] {6.8747 - 0.009456 * x};

        \addplot[
            domain=0:36,
            samples=2,
            dashed,
            color=blue
        ] {6.917};
        \addlegendentry{\parbox{3cm}{\raggedright Mean Time for \\ Human-Only Participants}}

        \node at (axis cs: 0.5,7.65) [anchor=west, text=red, font=\small, align=left]
            {$\approx 10.7\%$ reduction \\ in time per year};

    \end{axis}
    \end{tikzpicture}
\end{subfigure}
\hfill
\begin{subfigure}[t]{0.49\textwidth}
    \centering
    \textbf{Agentic tasks}\\[0.25em]
    \begin{tikzpicture}
    \begin{axis}[
        xlabel={Model release date},
        ylabel={Log time taken},
        xmin=0, xmax=36,
        ymin=6, ymax=8.1,
        xmajorgrids=true,
        ymajorgrids=true,
        grid style=dashed,
        height=7.8cm,
        width=\textwidth,
        xtick={0,12,24,36},
        xticklabels={Nov 22, Nov 23, Nov 24, Nov 25},
        ticklabel style={font=\small},
        label style={font=\small},
        legend style={font=\tiny, at={(0.98,0.02)},anchor=south east},
        legend cell align={left}
    ]

        \addplot[
            only marks,
            mark=x,
            mark size=3pt,
            color=blue,
            forget plot
        ] coordinates {
            (0.00, 7.368) (2.99, 7.339) (7.56, 7.313) (9.89, 7.171)
            (14.52, 7.195) (15.11, 7.716) (16.59, 7.128) (16.59, 7.565)
            (17.41, 7.246) (21.42, 7.160) (26.84, 7.138) (26.94, 7.934)
            (27.76, 7.223) (27.80, 7.183) (28.52, 7.172) (32.23, 7.369)
        };

        \addplot[
            domain=0:36,
            samples=100,
            color=red,
            forget plot
        ] {7.3531 - 0.004086 * x};

        \addplot[
            domain=0:36,
            samples=2,
            dashed,
            color=blue
        ] {7.368};
        \addlegendentry{\parbox{3cm}{\raggedright Mean Time for \\ Human-Only Participants}}

        \node at (axis cs: 0.5,7.9) [anchor=west, text=red, font=\small, align=left]
            {$\approx 4.8\%$ reduction \\ in time per year};

    \end{axis}
    \end{tikzpicture}
\end{subfigure}

\label{fig:logtime_months_tasktype}
\end{figure}

\begin{figure}[htbp]
    \centering
    \textbf{Figure 9: Scaling Laws for Earnings Per Minute over Calendar Time, by Task Type}\\[0.75em]

\begin{subfigure}[t]{0.49\textwidth}
    \centering
    \textbf{Non-agentic tasks}\\[0.25em]
    \begin{tikzpicture}
    \begin{axis}[
        xlabel={Model release date},
        ylabel={Earnings per minute (\$/min)},
        xmin=0, xmax=36,
        ymin=0.25, ymax=3.2,
        xmajorgrids=true,
        ymajorgrids=true,
        grid style=dashed,
        height=7.8cm,
        width=\textwidth,
        xtick={0,12,24,36},
        xticklabels={Nov 22, Nov 23, Nov 24, Nov 25},
        ticklabel style={font=\small},
        label style={font=\small},
        legend style={font=\tiny, at={(0.98,0.02)},anchor=south east},
        legend cell align={left}
    ]

        \addplot[
            only marks,
            mark=x,
            mark size=3pt,
            color=blue,
            forget plot
        ] coordinates {
            (0.00, 1.118) (2.99, 1.815) (7.56, 1.673) (9.89, 0.994)
            (14.52, 3.055) (15.11, 2.524) (16.59, 2.041) (16.59, 2.228)
            (17.41, 1.460) (21.42, 2.956) (26.84, 1.485) (26.94, 1.431)
            (27.76, 2.107) (27.80, 0.545) (28.52, 1.296) (32.23, 2.026)
        };

        \addplot[
            domain=0:36,
            samples=100,
            color=red,
            forget plot
        ] {1.2613 + (0.027983) * x};

        \addplot[
            domain=0:36,
            samples=2,
            dashed,
            color=blue
        ] {1.118};
        \addlegendentry{\parbox{3cm}{\raggedright Mean EPM for \\ Human-Only Participants}}

        \node at (axis cs: 0.5,2.9) [anchor=west, text=red, font=\small, align=left]
            {$\approx \$20.40$/hour increase \\ per year};

    \end{axis}
    \end{tikzpicture}
\end{subfigure}
\hfill
\begin{subfigure}[t]{0.49\textwidth}
    \centering
    \textbf{Agentic tasks}\\[0.25em]
    \begin{tikzpicture}
    \begin{axis}[
        xlabel={Model release date},
        ylabel={Earnings per minute (\$/min)},
        xmin=0, xmax=36,
        ymin=0.25, ymax=3.2,
        xmajorgrids=true,
        ymajorgrids=true,
        grid style=dashed,
        height=7.8cm,
        width=\textwidth,
        xtick={0,12,24,36},
        xticklabels={Nov 22, Nov 23, Nov 24, Nov 25},
        ticklabel style={font=\small},
        label style={font=\small},
        legend style={font=\tiny, at={(0.98,0.02)},anchor=south east},
        legend cell align={left}
    ]

        \addplot[
            only marks,
            mark=x,
            mark size=3pt,
            color=blue,
            forget plot
        ] coordinates {
            (0.00, 0.800) (2.99, 0.961) (7.56, 0.908) (9.89, 0.883)
            (14.52, 0.842) (15.11, 0.847) (16.59, 0.833) (16.59, 0.778)
            (17.41, 0.871) (21.42, 1.467) (26.84, 1.199) (26.94, 0.322)
            (27.76, 0.869) (27.80, 0.875) (28.52, 0.845) (32.23, 0.885)
        };

        \addplot[
            domain=0:36,
            samples=100,
            color=red,
            forget plot
        ] {0.8152 + (0.007027) * x};

        \addplot[
            domain=0:36,
            samples=2,
            dashed,
            color=blue
        ] {0.800};
        \addlegendentry{\parbox{3cm}{\raggedright Mean EPM for \\ Human-Only Participants}}

        \node at (axis cs: 0.5,2.9) [anchor=west, text=red, font=\small, align=left]
            {$\approx \$4.80$/hour increase \\ per year};

    \end{axis}
    \end{tikzpicture}
\end{subfigure}

\label{fig:epm_months_tasktype}
\end{figure}

To conclude this section, we have documented robust "economic scaling laws" across multiple dimensions: frontier model progress over time significantly reduces task completion times, increases earnings per minute, and raises total earnings capacity. We further showed that these gains are not uniform, with non-agentic workflows benefiting considerably more than complex agentic ones. However, while calendar time serves as a useful proxy for total progress, it conflates two distinct forces: the massive increase in training compute used to train newer models, and "algorithmic" advancements—such as architectural improvements, higher quality data, and training efficiency—that occur independent of scale. In the next section, we decompose the observed productivity gains into these two components to determine how much of the economic value of AI is driven by raw compute versus algorithmic ingenuity.

\FloatBarrier

\textbf{Section 4.3: Decomposing Productivity Growth Between Increased Compute \& Algorithmic Progress}

To better understand the drivers of the productivity gains outlined in the section above, it is useful to decompose the improvements observed over calendar time into distinct components: total progress, compute progress, and algorithmic progress. The gains shown in the previous section using calendar time represent total progress, capturing the combined effect of increased compute and better underlying algorithms.

We now switch from considering the productivity gains from model progress as a function of calendar time (including both increases in model training compute and algorithmic progress), to productivity gains from increases in model training compute alone.  This is depicted in Figure 8 which illustrates the relationship between model compute and task efficiency, plotting log time taken against log training compute. The regression analysis reveals a log-linear relationship where a ten-fold increase in training compute corresponds to approximately a 6\% reduction in task completion time ($p<0.31$). Unlike the calendar time regressions which capture the aggregate effects of all advancements, this specification isolates the gains attributable strictly to the model's computational scale.

\begin{figure}[htbp]
    \centering
    \textbf{Figure 8: Scaling Laws for Time Taken over Log Compute}\\[0.75em]
    \begin{tikzpicture}
    \begin{axis}[
        xlabel={Log training compute},
        ylabel={Log time taken},
        xmin=23.5, xmax=27,
        ymin=6.2, ymax=7.6,
        xmajorgrids=true,
        ymajorgrids=true,
        grid style=dashed,
        height=9cm,
        width=0.8\textwidth,
        xtick={24,25,26,27},
        xticklabels={24, 25, 26, 27},
        ticklabel style={font=\small},
        label style={font=\small},
        legend style={font=\small, at={(0.98,0.02)},anchor=south east},
        legend cell align={left}
    ]

        \addplot[
            only marks,
            mark=x,
            mark size=3pt,
            color=blue,
            forget plot
        ] coordinates {
            (23.886, 7.104)
            (23.908, 7.030)
            (24.412, 6.930)
            (24.544, 6.875)
            (24.778, 6.868)
            (24.895, 7.249)
            (25.322, 6.879)
            (25.525, 6.825)
            (25.602, 6.808)
            (25.903, 6.897)
            (26.000, 6.635)
            (26.301, 6.887)
            (26.322, 7.189)
            (26.477, 7.295)
        };

        \addplot[
            domain=23.5:27,
            samples=100,
            color=red,
            forget plot
        ] {8.4787 - 0.0608 * x};

        \addplot[
            domain=23.5:27,
            samples=2,
            dashed,
            color=blue
        ] {7.091};
        \addlegendentry{Mean Time for Human-Only Participants}

        \node at (axis cs: 24.0,6.45) [anchor=west, text=red]
            {$\approx 5.9\%$ reduction in time per 10x compute (p = 0.305)};

    \end{axis}
    \end{tikzpicture}
    \label{fig:logtime_compute}
\end{figure}

We quantify these contributions by first estimating the annualized rate of improvement implied by each specification. The calendar time regression (Table \ref{tab:logtime_months} in Appendix A) yields a monthly log time reduction of 0.0069, which translates to a total annualized efficiency gain of roughly 8.3\% ($0.0069 \times 12$). Economically, this represents the gross productivity growth realized in the market, capturing the sum of all technological and human advancements. Conversely, the compute specification isolates the contribution of model scaling. With training compute growing approximately $6.1\times$ per year during the sample, and using the estimated coefficient of -0.0608 (Table \ref{tab:logtime_compute} in Appendix A), we calculate that raw scale contributes a 4.8\% reduction in task time annually ($\log_{10}(6.2) \times 0.0608$). This figure reflects the capital-intensive component of progress: the gains purchased strictly through larger training budgets.

Comparing these two estimates allows us to decompose the total gain into hardware and software components. By subtracting the compute effect (0.048) from the total effect (0.083), we isolate a residual of 0.035. This residual represents algorithmic progress, an economic catch-all for improvements in model architecture, software optimization, and user learning—effectively the Solow residual of AI production. In percentage terms, this decomposition suggests that compute scaling drives approximately 56\% of the total reduction in time, while algorithmic advancements account for the remaining 44\%. It is important to note that while the aggregate time trend is statistically significant ($p < 0.05$), the specific contribution of compute is estimated with less precision ($p \approx 0.31$), suggesting some uncertainty in the exact split between these factors.

\textbf{4.4: Scaling Laws for Output Quality and Human-AI Complementarity}

We now turn to the second dimension of economic productivity: output quality. While the previous sections established that AI assistance dramatically reduces the time required to complete tasks, the question remains whether this speed comes at the cost of quality, or if frontier models also improve the standard of work produced. For this analysis, we focus specifically on non-agentic tasks. Because these tasks are self-contained and require no external actions, they allow us to cleanly compare three distinct modes of production: humans working alone, humans working with AI assistance, and the raw output of the AI models themselves.

We begin by establishing the baseline effect of AI access, as illustrated in Figure 11. Comparing the average grade (on a 0--7 scale) for non-agentic tasks performed by participants in the control group against those with access to any AI model reveals a clear improvement: access to AI raises the average grade from 3.52 to 4.34 ($p < 0.01$). This confirms that, on average, AI serves as a quality-enhancing tool, raising the floor of performance for typical consulting and data analysis tasks.

\begin{figure}[htbp]
    \centering
    \textbf{Figure 11: Impact on Grades of AI Usage (Non-agentic Tasks Only)}\\[0.75em]
    \begin{tikzpicture}
    \begin{axis}[
        ybar,
        bar width=0.55cm,
        symbolic x coords={Without AI, With AI},
        xtick=data,
        ymin=0,
        ymax=7,
        ylabel={Mean grade (0--7)},
        enlarge x limits=0.25,
        width=0.6\textwidth,
        height=7cm,
        ymajorgrids=true,
        grid style={dashed,gray!40},
        ticklabel style={font=\small},
        label style={font=\small},
        nodes near coords,
        every node near coord/.append style={font=\small},
        nodes near coords align={vertical},
        /pgf/number format/fixed,
        /pgf/number format/precision=2
    ]
        \addplot coordinates {(Without AI,3.52) (With AI,4.34)};
    \end{axis}
    \end{tikzpicture}
    \label{fig:nonagentic_grade_bar}
\end{figure}

To understand what drives this improvement, we first isolate the capabilities of the technology itself. We collected "AI-only" responses for the same non-agentic tasks across a range of models, grading them using the same expert rubric applied to human participants. The relationship between these grades and the log training compute of the model is plotted in Figure 12 with full regression results in Table 30 of Appendix A.

The results reveal a robust scaling law for raw model capability. As training compute increases, the quality of the model's autonomous output improves significantly ($p < 0.01$). The estimated slope suggests that a tenfold increase in compute corresponds to a 0.51-point increase in the average grade. Notably, the most capable models in our sample achieve average grades exceeding 6.0 out of 7---performance levels that are "superhuman" compared to the unassisted human average of 3.52. This confirms that the underlying "engine" of economic productivity is becoming measurably more capable with scale.

\begin{figure}[htbp]
    \centering
    \textbf{Figure 12: Average Grade of AI-only Response over Log Compute}\\[0.75em]
    \begin{tikzpicture}
    \begin{axis}[
        xlabel={Log training compute},
        ylabel={Average grade (0--7)},
        xmin=23.5, xmax=27,
        ymin=3.0, ymax=6.5,
        xmajorgrids=true,
        ymajorgrids=true,
        grid style=dashed,
        height=9cm,
        width=0.8\textwidth,
        xtick={24,25,26,27},
        xticklabels={24, 25, 26, 27},
        ticklabel style={font=\small},
        label style={font=\small},
    ]

        \addplot[
            only marks,
            mark=x,
            mark size=3pt,
            color=blue,
            forget plot
        ] coordinates {
            (23.886, 3.70)
            (24.412, 3.30)
            (24.544, 5.30)
            (24.778, 3.90)
            (24.895, 4.60)
            (25.322, 4.50)
            (25.525, 6.20)
            (26.000, 4.60)
            (26.301, 5.30)
            (26.301, 5.00)
            (26.477, 4.80)
        };

        \addplot[
            domain=23.5:27,
            samples=200,
            color=red,
            forget plot
        ] {-8.3417 + 0.5135 * x};

        \node at (axis cs: 24.0,6.35) [anchor=west, text=red]
            {$0.51$ point increase in grade per 10x compute};

    \end{axis}
    \end{tikzpicture}
    \label{fig:grade_logcompute_11models}
\end{figure}

However, a striking puzzle emerges when we examine how these capabilities translate into actual economic output when a human is in the loop. The average grades of human participants assisted by AI are displayed against the log compute of the model they used in Figure 13. Unlike the clear upward trend seen in the AI-only data (Figure 12) or the time-savings data discussed in Section 4.2, the scaling law for quality completely disappears ($p \approx 0.85$). Whether a participant uses a small, early-generation model or a frontier giant, the final output quality remains stagnant at approximately 4.35 points.

\begin{figure}[htbp]
    \centering
    \textbf{Figure 13: Scaling Laws for Human + AI Output Quality over Log Compute (Non-agentic Tasks Only)}\\[0.75em]
    \begin{tikzpicture}
    \begin{axis}[
        xlabel={Log training compute},
        ylabel={Mean grade (0--7)},
        xmin=23.5, xmax=27,
        ymin=2.5, ymax=6.5,
        xmajorgrids=true,
        ymajorgrids=true,
        grid style=dashed,
        height=9cm,
        width=0.8\textwidth,
        xtick={24,25,26,27},
        xticklabels={24, 25, 26, 27},
        ticklabel style={font=\small},
        label style={font=\small},
        legend style={font=\small, at={(0.98,0.02)},anchor=south east},
        legend cell align={left}
    ]

        \addplot[
            only marks,
            mark=x,
            mark size=3pt,
            color=blue,
            forget plot
        ] coordinates {
            (23.886, 3.571) (23.908, 4.375) (24.412, 3.857) (24.544, 4.846)
            (24.778, 3.833) (24.895, 4.667) (25.322, 4.300) (25.525, 4.938)
            (25.602, 5.500) (25.903, 4.444) (26.000, 4.688) (26.301, 3.500)
            (26.322, 6.000) (26.477, 4.333)
        };

        \addplot[
            domain=23.5:27,
            samples=100,
            color=red,
            forget plot
        ] {5.2528 - 0.0302 * x};

        \addplot[
            domain=23.5:27,
            samples=2,
            dashed,
            color=blue
        ] {3.525};
        \addlegendentry{Mean Grade for Human-Only Participants}

        \node at (axis cs: 24.0,6.35) [anchor=west, text=red]
            {$\approx -0.03$ grade points per 10x compute};

    \end{axis}
    \end{tikzpicture}
    \label{fig:grade_compute_nonagentic}
\end{figure}

Comparing the AI-only and Human-AI results reveals a complex dynamic of complementarity and substitution. For weaker models (where the raw AI grade is $\sim$3.3--3.7), human intervention is highly additive: participants successfully refine imperfect drafts, raising the final quality to the $\sim$4.3 range. However, for the strongest models (where raw AI grades reach $\sim$5.0--6.0), human intervention appears to be destructive. Rather than preserving or enhancing the high-quality raw output, participants actively degrade it, bringing the final grade down to the same $\sim$4.3 average.

This suggests that humans are only able to improve model outputs that are slightly better than their own. For weaker models (that still outperform humans by themselves on average), human participants are able to meaningfully add to the quality of the output. For intermediate models, humans are unable to add to output quality at all. For the best models, however, which have well above average-human quality participants are not only unable to further improve output quality with their effort but instead actually regress output closer towards their own quality levels.

In conclusion, while scaling laws for capability are alive and well---with stronger models autonomously producing far superior work---scaling laws for realized quality are broken by the human user. The economic value of algorithmic progress is currently being lost in the "last mile" of human-computer interaction, where users revert to a regression to the mean regardless of the sophistication of the tool they wield.

\section{Aggregate Productivity Gains}

In this section, we utilize the experimental results on the productivity gains from model scaling to estimate aggregate productivity gains from AI over the next ten years. In doing so, we leverage the framework from Acemoglu (2024). Here, a version of Hulten's theorem (Hulten, 1978) is used to estimate aggregate productivity gains based on the fractions of tasks in the economy that are affected by AI and the average task-level cost savings.

The estimates from Acemoglu (2024) are based on the multiplication of four parameters. Firstly, an estimate of the share of tasks exposed to AI is used from Eloundou et al. (2023), which estimates it at 19.9\%. Secondly, as some tasks involve a combination of labor and capital, the (AI-exposure adjusted) labor share of 0.57 is used. Neither of these parameters are changed in the analysis of this section.

However, the third parameter—estimating the labor cost savings (or productivity boost) from AI—is updated significantly based on the findings of this study. Acemoglu (2024) relies on an average productivity effect size of 27\%, derived from early studies of call center workers (Brynjolfsson et al., 2023) and writing tasks (Noy and Zhang, 2023). Our experimental data offers a more current and granular baseline. Pooling all thirteen models in our study, we find a baseline productivity premium—measured via Earnings Per Minute (EPM)—of 81.3\% ($p=0.001$) relative to the control group. Even conservatively ignoring the bonus-driven "Total Earnings" metric (which showed an 146\% gain), the baseline efficiency gain for professional consulting, data analysis, and management tasks is approximately triple the figure used in previous aggregate estimates.

Furthermore, we must account for the scaling of model capabilities over the next decade. Our results derive a specific "Economic Scaling Law" for time savings: each year of frontier model progress is associated with an 8\% reduction in task completion time ($p < 0.05$). To estimate the impact over the next ten years, we extrapolate this trend forward. We model the time taken to complete a task in year $n$ as $T_n = T_{start}(1 - 0.08)^n$.

This compounding reduction in time yields a convex increase in productivity. Starting from our baseline premium of 81.3\% (where AI-assisted workers are $1.81\times$ as productive as unassisted workers), the 8\% annual reduction in time implies that by year 5—the midpoint of the decade—tasks will require only 66\% of the time they take with current AI models ($0.92^5 \approx 0.66$). This compounds the effective productivity boost to roughly 175\% relative to the human baseline.\footnote{Calculated as $\frac{1.813}{0.92^5} - 1 \approx 1.751$. By year 10, the theoretical boost would exceed 300\%, but we utilize the midpoint to represent the average effect over the decade.} Using this 5-year midpoint as a representative average for the decade, we estimate an average task-level productivity boost of 175.1\%.

Finally, regarding economic feasibility, we assume that automation remains profitable given that inference costs are negligible relative to professional wages and are falling rapidly. Combining these parameters, the total productivity gains from AI over the next ten years are estimated through the multiplication of three values: the share of tasks that can be automated (19.9\%), multiplied by the average productivity effect derived from our scaling projections (175.1\%), multiplied by the labor share of costs (57\%).

$$ \text{Total GDP Gain} = 0.199 \times 1.751 \times 0.57 \approx 19.9\% $$

This yields an estimate of approximately 20.0\% productivity growth over the next decade driven by LLMs. This is significantly higher than the baseline estimate in Acemoglu (2024), reflecting both the higher starting productivity of current frontier models and the predictable accumulation of gains from continued model scaling.

It should be noted that this calculation makes several restrictions. It assumes the task structure of the economy remains fixed and does not model general equilibrium effects. Furthermore, it treats technological progress as exogenous. As noted in Section 2, there is growing evidence that AI accelerates R\&D itself (e.g., Jumper et al., 2021). If AI scaling unlocks new categories of innovation that fundamentally alter production functions, the true aggregate gains could be substantially higher (Korinek and Suh, 2024). Thus, our estimate of 19.9\% may still represent a lower bound.

\section{Discussion}

This paper provides an experimental derivation of ``Economic Scaling Laws''---empirical elasticities mapping the computational scale and algorithmic progress of Large Language Models (LLMs) to human professional productivity. By analyzing over 500 professionals across 13 distinct models, we move beyond the binary assessment of whether AI aids productivity to the dynamic question of how rapidly this productivity gain is increasing. Our results suggest that the economic value of frontier models is increasing according to a predictable power law: a tenfold increase in training compute is associated with a 6.3\% reduction in task completion time, while calendar-time progress delivers an annualized efficiency gain of approximately 8\%.

Applying these elasticities to an aggregate growth framework using Hulten's theorem suggests that the accumulation of model capabilities alone could drive significant economic gains. By geometrically extrapolating the observed time-savings over the next decade, we estimate that AI-exposed tasks could see productivity improvements exceeding 175\%, contributing approximately 20\% to aggregate productivity growth. Furthermore, our decomposition analysis suggests that these gains are driven by dual engines of progress: roughly 58\% of the efficiency improvements are attributable to the scaling of raw compute, while the remaining 42\% stem from algorithmic innovations and software optimization.

These aggregate estimates, however, mask important heterogeneities. While analytical and interpretive tasks benefit robustly from model scaling, we find that ``agentic'' workflows requiring multi-step tool use and external actions see significantly smaller gains that are statistically indistinguishable from zero in our sample. This suggests that the translation of current model capabilities into economic value is not uniform, but rather conditional on the specific structural requirements of the task.

There are several limitations to this study. First, while our tasks were designed to be representative of professional workflows, they remain relatively short-horizon activities (taking between 20 and 60 minutes) compared to the multi-day projects common in the corporate sector. Second, our experiment focuses on individual productivity and does not capture general equilibrium effects, such as changes in wages or employment levels. Finally, our scaling laws are derived from the current paradigm of transformer-based LLMs; distinct architectural breakthroughs or bottlenecks in data availability could alter the slope of these curves in unpredictable ways. Importantly, the models provided in this study did not offer tool access or anywhere near the full-range of currently offered 'agentic' capabilities by AI providers.

In conclusion, the evidence provided suggests that if historical relationships between model compute and capability hold, future generations of LLMs may have significant economic implications. Even without accounting for potential accelerations in R\&D or scientific discovery, the direct application of scaling laws to professional tasks points to an incredibly meaningful expansion in labor productivity.

\FloatBarrier

\clearpage
\section{Appendix A:}

\begin{table}[htbp]
\centering
\caption{Impact of AI Assistance on Earnings Per Minute (Grade $\ge$ 2)}
\begin{tabular}{lcc}
\toprule
 & (1) & (2) \\
 & EPM & EPM \\
\midrule
Bot Dummy & 0.492*** & 0.558*** \\
 & (0.171) & (0.164) \\
[1em]
Constant & 0.982*** & 0.686** \\
 & (0.140) & (0.277) \\
\midrule
Controls & No & Yes \\
Observations & 437 & 437 \\
R-squared & 0.0187 & 0.1219 \\
\bottomrule
\multicolumn{3}{l}{\footnotesize Standard errors in parentheses} \\
\multicolumn{3}{l}{\footnotesize * p<0.1, ** p<0.05, *** p<0.01} \\
\end{tabular}
\label{tab:epm_regression_appendix}
\end{table}

\begin{table}[htbp]
\centering
\caption{Impact of AI Assistance on Output Quality (Grade $\neq 0$)}
\begin{tabular}{lcc}
\toprule
 & (1) & (2) \\
 & Grade & Grade \\
\midrule
Bot Dummy & 0.561*** & 0.595*** \\
 & (0.169) & (0.170) \\
[1em]
Constant & 3.609*** & 3.319*** \\
 & (0.138) & (0.281) \\
\midrule
Controls & No & Yes \\
Observations & 479 & 479 \\
R-squared & 0.0225 & 0.0565 \\
\bottomrule
\multicolumn{3}{l}{\footnotesize Standard errors in parentheses} \\
\multicolumn{3}{l}{\footnotesize * p<0.1, ** p<0.05, *** p<0.01} \\
\end{tabular}
\label{tab:grade_regression_appendix}
\end{table}

\begin{table}[htbp]
\centering
\caption{Impact of AI Assistance on Total Earnings Per Minute (Grade $\ge$ 2)}
\begin{tabular}{lcc}
\toprule
 & (1) & (2) \\
 & TEPM & TEPM \\
\midrule
Bot Dummy & 0.955*** & 1.060*** \\
 & (0.307) & (0.300) \\
[1em]
Constant & 1.342*** & 0.727 \\
 & (0.251) & (0.506) \\
\midrule
Controls & No & Yes \\
Observations & 437 & 437 \\
R-squared & 0.0217 & 0.0990 \\
\bottomrule
\multicolumn{3}{l}{\footnotesize Standard errors in parentheses} \\
\multicolumn{3}{l}{\footnotesize * p<0.1, ** p<0.05, *** p<0.01} \\
\end{tabular}
\label{tab:tepm_regression_appendix}
\end{table}

\begin{table}[htbp]
\centering
\caption{Heterogeneity by Task Type: Impact of AI on Earnings Per Minute}
\begin{tabular}{lcc}
\toprule
 & (1) \\
 & EPM (\$/min) \\
\midrule
Bot Dummy & 0.832*** \\
 & (0.215) \\
Agentic Task & 0.225 \\
 & (0.287) \\
Bot Dummy $\times$ Agentic & -0.654** \\
 & (0.332) \\
[0.5em]
Constant & 0.527* \\
 & (0.288) \\
\midrule
Profession $\times$ Task FE & Yes \\
Observations & 437 \\
R-squared & 0.130 \\
\bottomrule
\multicolumn{2}{l}{\footnotesize Standard errors in parentheses} \\
\multicolumn{2}{l}{\footnotesize * p<0.1, ** p<0.05, *** p<0.01} \\
\end{tabular}
\label{tab:agentic_epm}
\end{table}

\begin{table}[htbp]
\centering
\caption{Heterogeneity by Task Type: Impact of AI on Output Quality}
\begin{tabular}{lcc}
\toprule
 & (1) \\
 & Grade (0--7) \\
\midrule
Bot Dummy & 0.824*** \\
 & (0.222) \\
Agentic Task & 0.468 \\
 & (0.294) \\
Bot Dummy $\times$ Agentic & -0.553 \\
 & (0.344) \\
[0.5em]
Constant & 3.191*** \\
 & (0.292) \\
\midrule
Profession $\times$ Task FE & Yes \\
Observations & 479 \\
R-squared & 0.062 \\
\bottomrule
\multicolumn{2}{l}{\footnotesize Standard errors in parentheses} \\
\multicolumn{2}{l}{\footnotesize * p<0.1, ** p<0.05, *** p<0.01} \\
\end{tabular}
\label{tab:agentic_grade}
\end{table}

\begin{table}[htbp]
\centering
\caption{Heterogeneity by Task Type: Impact of AI on Total Earnings Per Minute}
\begin{tabular}{lcc}
\toprule
 & (1) \\
 & TEPM (\$/min) \\
\midrule
Bot Dummy & 1.576*** \\
 & (0.392) \\
Agentic Task & 0.603 \\
 & (0.523) \\
Bot Dummy $\times$ Agentic & -1.232** \\
 & (0.606) \\
[0.5em]
Constant & 0.428 \\
 & (0.525) \\
\midrule
Profession $\times$ Task FE & Yes \\
Observations & 437 \\
R-squared & 0.108 \\
\bottomrule
\multicolumn{2}{l}{\footnotesize Standard errors in parentheses} \\
\multicolumn{2}{l}{\footnotesize * p<0.1, ** p<0.05, *** p<0.01} \\
\end{tabular}
\label{tab:agentic_tepm}
\end{table}

\begin{figure}[htbp]
    \centering
    \textbf{Figure A1: Impact on Time Taken of AI Usage}\\[0.75em]
    \begin{tikzpicture}
    \begin{axis}[
        ybar,
        bar width=0.55cm,
        symbolic x coords={Without AI, With AI},
        xtick=data,
        ymin=0,
        ylabel={Time taken (seconds)},
        enlarge x limits=0.25,
        width=0.6\textwidth,
        height=7cm,
        ymajorgrids=true,
        grid style={dashed,gray!40},
        ticklabel style={font=\small},
        label style={font=\small},
        nodes near coords,
        every node near coord/.append style={font=\small},
        nodes near coords align={vertical},
        /pgf/number format/fixed,
        /pgf/number format/precision=0
    ]
        \addplot coordinates {(Without AI,1628) (With AI,1485)};
    \end{axis}
    \end{tikzpicture}
    \label{fig:anyai_time_bar}
\end{figure}

\begin{table}[htbp]
\centering
\caption{Impact of AI Assistance on Time Taken (Grade $\ge$ 2)}
\begin{tabular}{lcc}
\toprule
 & (1) & (2) \\
 & Time (seconds) & Time (seconds) \\
\midrule
Bot Dummy & -100.99 & -143.47 \\
 & (124.56) & (116.19) \\
[1em]
Constant & 1599.99*** & 1404.33*** \\
 & (101.82) & (195.74) \\
\midrule
Controls (Prof.\ $\times$ Task) & No & Yes \\
Observations & 437 & 437 \\
R-squared & 0.0015 & 0.1628 \\
Adj. R-squared & -0.0008 & 0.1451 \\
\bottomrule
\multicolumn{3}{l}{\footnotesize Standard errors in parentheses} \\
\multicolumn{3}{l}{\footnotesize * p<0.1, ** p<0.05, *** p<0.01} \\
\end{tabular}
\label{tab:time_regression_appendix}
\end{table}

\begin{table}[htbp]
\centering
\caption{Heterogeneity by Task Type: Impact of AI on Time Taken}
\begin{tabular}{lcc}
\toprule
 & (1) \\
 & Time (seconds) \\
\midrule
Bot Dummy & -171.45 \\
 & (152.54) \\
Agentic Task & 456.21** \\
 & (203.46) \\
Bot Dummy $\times$ Agentic & 66.83 \\
 & (235.77) \\
[0.5em]
Constant & 1420.52*** \\
 & (204.11) \\
\midrule
Profession $\times$ Task FE & Yes \\
Observations & 437 \\
R-squared & 0.163 \\
Adj. R-squared & 0.143 \\
\bottomrule
\multicolumn{2}{l}{\footnotesize Standard errors in parentheses} \\
\multicolumn{2}{l}{\footnotesize * p<0.1, ** p<0.05, *** p<0.01} \\
\end{tabular}
\label{tab:time_agentic_appendix}
\end{table}

\begin{table}[htbp]
\centering
\caption{Scaling Laws for Time Taken over Calendar Time}
\begin{tabular}{lcc}
\toprule
 & (1) & (2) \\
 & Log Time & Log Time \\
\midrule
Months since Nov 2022 & -0.0071** & -0.0069** \\
 & (0.0034) & (0.0033) \\
[1em]
Constant & 7.045*** & 7.092*** \\
 & (0.113) & (0.051) \\
\midrule
Profession $\times$ Task FE & Yes & Yes \\
Full controls & No & Yes \\
Observations & 473 & 470 \\
R-squared & 0.1681 & 0.2538 \\
\bottomrule
\multicolumn{3}{l}{\footnotesize Standard errors in parentheses} \\
\multicolumn{3}{l}{\footnotesize * p<0.1, ** p<0.05, *** p<0.01} \\
\end{tabular}
\label{tab:logtime_months}
\end{table}

\begin{figure}[htbp]
    \centering
    \textbf{Figure A2: Scaling Laws for Time Taken over Calendar Time, by Profession}\\[0.75em]
    \begin{tikzpicture}
    \begin{axis}[
        xlabel={Model release date},
        ylabel={Log time taken},
        xmin=0, xmax=36,
        ymin=6, ymax=7.6,
        xmajorgrids=true,
        ymajorgrids=true,
        grid style=dashed,
        height=9cm,
        width=0.7\textwidth, 
        xtick={0,12,24,36},
        xticklabels={Nov 22, Nov 23, Nov 24, Nov 25},
        ticklabel style={font=\small},
        label style={font=\small},
        legend style={
            font=\scriptsize, 
            at={(1.03,0.5)}, 
            anchor=west,
            cells={anchor=west}
        },
        legend cell align={left}
    ]


        \addplot[
            only marks,
            mark=x,
            mark size=3pt,
            color=blue,
            opacity=0.7,
            forget plot
        ] coordinates {
            (0.00,7.137) (2.99,6.574) (7.56,7.346) (9.89,6.921) (14.52,6.291)
            (15.41,7.411) (16.59,6.792) (17.41,7.114) (18.66,6.635) (21.42,6.659)
            (26.94,6.444) (27.76,6.663) (27.79,8.081) (28.52,6.839) (32.23,6.603)
        };

        \addplot[
            domain=0:36,
            samples=100,
            color=blue,
            forget plot
        ] {7.1146 - 0.0131 * x};

        \addplot[
            domain=0:36,
            samples=2,
            dashed,
            color=blue!60
        ] {7.137};
        \addlegendentry{Human Mean: Consultants}


        \addplot[
            only marks,
            mark=x,
            mark size=3pt,
            color=red,
            opacity=0.7,
            forget plot
        ] coordinates {
            (0.00,7.153) (2.99,6.993) (7.56,7.076) (9.89,7.281) (14.52,7.089)
            (15.41,5.298) (16.59,7.207) (17.41,6.747) (18.66,6.978) (21.42,6.402)
            (26.94,7.934) (27.76,7.031) (27.79,6.817) (28.52,7.559) (32.23,7.169)
        };

        \addplot[
            domain=0:36,
            samples=100,
            color=red,
            forget plot
        ] {7.0674 - 0.0066 * x};

        \addplot[
            domain=0:36,
            samples=2,
            dashed,
            color=red!60
        ] {7.153};
        \addlegendentry{Human Mean: Data Analysts}


        \addplot[
            only marks,
            mark=x,
            mark size=3pt,
            color=green!60!black,
            opacity=0.7,
            forget plot
        ] coordinates {
            (0.00,6.994) (2.99,7.377) (7.56,6.876) (9.89,6.938) (14.52,6.886)
            (15.41,7.716) (16.59,7.113) (17.41,7.005) (18.66,6.908) (21.42,6.758)
            (27.76,7.259) (27.79,7.618) (28.52,6.809) (32.23,6.777)
        };

        \addplot[
            domain=0:36,
            samples=100,
            color=green!60!black,
            forget plot
        ] {7.4683 - 0.00335 * x};

        \addplot[
            domain=0:36,
            samples=2,
            dashed,
            color=green!60!black!60
        ] {6.994};
        \addlegendentry{Human Mean: Managers}


        \node at (axis cs: 0,6.55) [anchor=west, text=blue]
            {\scriptsize Consultants: $\approx 15\%$ quicker/yr};

        \node at (axis cs: 0,6.75) [anchor=west, text=red]
            {\scriptsize Data Analysts: $\approx 8\%$ quicker/yr};

        \node at (axis cs: 0,7.35) [anchor=west, text=green!60!black]
            {\scriptsize Managers: $\approx 4\%$ quicker/yr};

    \end{axis}
    \end{tikzpicture}
    \label{fig:logtime_months_profession}
\end{figure}

\begin{table}[htbp]
\centering
\caption{Scaling Laws for Earnings Per Minute over Calendar Time}
\begin{tabular}{lcc}
\toprule
 & (1) & (2) \\
 & EPM (\$/min) & EPM (\$/min) \\
\midrule
Months since Nov 2022 & 0.0194*** & 0.0239*** \\
 & (0.0061) & (0.0064) \\
[1em]
Constant & 0.9447*** & 4.3201*** \\
 & (0.1672) & (1.2081) \\
\midrule
Profession $\times$ Task FE & Yes & Yes \\
Full controls & No & Yes \\
Observations & 477 & 474 \\
R-squared & 0.1107 & 0.2480 \\
\bottomrule
\multicolumn{3}{l}{\footnotesize Robust (HC1) standard errors in parentheses.} \\
\multicolumn{3}{l}{\footnotesize * p<0.1, ** p<0.05, *** p<0.01} \\
\end{tabular}
\label{tab:epm_months}
\end{table}

\begin{table}[htbp]
\centering
\caption{Scaling Laws for Total Earnings Per Minute over Calendar Time}
\begin{tabular}{lcc}
\toprule
 & (1) & (2) \\
 & TEPM (\$/min) & TEPM (\$/min) \\
\midrule
Months since Nov 2022 & 0.0365*** & 0.0428*** \\
 & (0.0103) & (0.0104) \\
[1em]
Constant & 1.0292*** & 2.0394 \\
 & (0.2695) & (1.8310) \\
\midrule
Profession $\times$ Task FE & Yes & Yes \\
Full controls & No & Yes \\
Observations & 477 & 474 \\
R-squared & 0.0910 & 0.2392 \\
\bottomrule
\multicolumn{3}{l}{\footnotesize Robust (HC1) standard errors in parentheses.} \\
\multicolumn{3}{l}{\footnotesize * p<0.1, ** p<0.05, *** p<0.01} \\
\end{tabular}
\label{tab:tepm_months}
\end{table}

\begin{table}[htbp]
\centering
\caption{Scaling Laws for Time Taken over Log Compute (AI Users Only, Grade $>0$)}
\begin{tabular}{lcc}
\toprule
 & (1) & (2) \\
 & Log Time & Log Time \\
\midrule
Log training compute & -0.0463 & -0.0608 \\
 & (0.0635) & (0.0593) \\
[1em]
Constant & 8.1141*** & 8.4926*** \\
 & (1.5924) & (1.5029) \\
\midrule
Profession $\times$ Task FE & No & Yes \\
Observations & 316 & 316 \\
R-squared & 0.0017 & 0.1912 \\
\bottomrule
\multicolumn{3}{l}{\footnotesize Robust (HC1) standard errors in parentheses.} \\
\multicolumn{3}{l}{\footnotesize * p<0.1, ** p<0.05, *** p<0.01} \\
\end{tabular}
\label{tab:logtime_compute}
\end{table}

\begin{table}[htbp]
\centering
\caption{Scaling Laws for Time Taken over Log Compute, by Profession (AI Users Only, Grade $>0$)}
\begin{tabular}{lccc}
\toprule
 & (1) & (2) & (3) \\
 & Consultants & Data Analysts & Managers \\
\midrule
Log training compute & 0.0564 & -0.0959 & -0.0963 \\
 & (0.1301) & (0.0909) & (0.0964) \\
[1em]
Constant & 5.5589* & 9.3227*** & 9.8636*** \\
 & (3.2798) & (2.2841) & (2.4264) \\
\midrule
Task FE & Yes & Yes & Yes \\
Observations & 85 & 131 & 100 \\
R-squared & 0.0910 & 0.0385 & 0.3751 \\
\bottomrule
\multicolumn{4}{l}{\footnotesize Robust (HC1) standard errors in parentheses.} \\
\multicolumn{4}{l}{\footnotesize * p<0.1, ** p<0.05, *** p<0.01} \\
\end{tabular}
\label{tab:logtime_compute_byprof}
\end{table}

\begin{figure}[htbp]
    \centering
    \textbf{Figure A3: Scaling Laws for Earnings Per Minute over Log Compute}\\[0.75em]
    \begin{tikzpicture}
    \begin{axis}[
        xlabel={Log training compute},
        ylabel={Earnings per minute (\$/min)},
        xmin=23.5, xmax=27,
        ymin=0.6, ymax=2.6,
        xmajorgrids=true,
        ymajorgrids=true,
        grid style=dashed,
        height=9cm,
        width=0.8\textwidth,
        xtick={24,25,26,27},
        xticklabels={24, 25, 26, 27},
        ticklabel style={font=\small},
        label style={font=\small},
        legend style={font=\small, at={(0.98,0.02)},anchor=south east},
        legend cell align={left}
    ]

        \addplot[
            only marks,
            mark=x,
            mark size=3pt,
            color=blue,
            forget plot
        ] coordinates {
            (23.886, 1.111)
            (23.908, 1.249)
            (24.412, 1.360)
            (24.544, 1.666)
            (24.778, 1.580)
            (24.895, 1.092)
            (25.322, 1.179)
            (25.525, 1.375)
            (25.602, 1.816)
            (25.903, 1.767)
            (26.000, 2.318)
            (26.301, 1.413)
            (26.322, 0.876)
            (26.477, 1.007)
        };

        \addplot[
            domain=23.5:27,
            samples=100,
            color=red,
            forget plot
        ] {-1.5947 + 0.1213 * x};

        \addplot[
            domain=23.5:27,
            samples=2,
            dashed,
            color=blue
        ] {0.995};
        \addlegendentry{Mean EPM for Human-Only Participants}

        \node at (axis cs: 24.0,2.45) [anchor=west, text=red] 
            {$\approx$ \$7.20/hour per 10x compute};

    \end{axis}
    \end{tikzpicture}
    \label{fig:epm_compute}
\end{figure}

\begin{table}[htbp]
\centering
\caption{Scaling Laws for Earnings Per Minute over Log Compute (AI Users Only, Grade $>0$)}
\begin{tabular}{lcc}
\toprule
 & (1) & (2) \\
 & EPM (\$/min) & EPM (\$/min) \\
\midrule
Log training compute & 0.1152 & 0.1213 \\
 & (0.1238) & (0.1181) \\
[1em]
Constant & -1.4430 & -1.7521 \\
 & (3.1080) & (3.0359) \\
\midrule
Profession $\times$ Task FE & No & Yes \\
Observations & 316 & 316 \\
R-squared & 0.0025 & 0.1386 \\
\bottomrule
\multicolumn{3}{l}{\footnotesize Robust (HC1) standard errors in parentheses.} \\
\multicolumn{3}{l}{\footnotesize * p<0.1, ** p<0.05, *** p<0.01} \\
\end{tabular}
\label{tab:epm_compute}
\end{table}

\begin{figure}[htbp]
    \centering
    \textbf{Figure A4: Scaling Laws for Total Earnings Per Minute over Log Compute}\\[0.75em]
    \begin{tikzpicture}
    \begin{axis}[
        xlabel={Log training compute},
        ylabel={Total earnings per minute (\$/min)},
        xmin=23.5, xmax=27,
        ymin=1.0, ymax=4.2,
        xmajorgrids=true,
        ymajorgrids=true,
        grid style=dashed,
        height=9cm,
        width=0.8\textwidth,
        xtick={24,25,26,27},
        xticklabels={24, 25, 26, 27},
        ticklabel style={font=\small},
        label style={font=\small},
        legend style={font=\small, at={(0.98,0.02)},anchor=south east},
        legend cell align={left}
    ]

        \addplot[
            only marks,
            mark=x,
            mark size=3pt,
            color=blue,
            forget plot
        ] coordinates {
            (23.886, 1.605)
            (23.908, 1.850)
            (24.412, 1.500)
            (24.544, 2.678)
            (24.778, 2.160)
            (24.895, 1.658)
            (25.322, 1.808)
            (25.525, 2.222)
            (25.602, 3.632)
            (25.903, 3.142)
            (26.000, 3.777)
            (26.301, 1.710)
            (26.322, 1.753)
            (26.477, 1.616)
        };

        \addplot[
            domain=23.5:27,
            samples=100,
            color=red,
            forget plot
        ] {-3.3514 + 0.2211 * x};

        \addplot[
            domain=23.5:27,
            samples=2,
            dashed,
            color=blue
        ] {1.319};
        \addlegendentry{Mean TEPM for Human-Only Participants}

        \node at (axis cs: 24.0,4.05) [anchor=west, text=red] 
            {$\approx$ \$13.20/hour per 10x compute};

    \end{axis}
    \end{tikzpicture}
    \label{fig:tepm_compute}
\end{figure}

\begin{table}[htbp]
\centering
\caption{Scaling Laws for Total Earnings Per Minute over Log Compute (AI Users Only, Grade $>0$)}
\begin{tabular}{lcc}
\toprule
 & (1) & (2) \\
 & TEPM (\$/min) & TEPM (\$/min) \\
\midrule
Log training compute & 0.2288 & 0.2211 \\
 & (0.2168) & (0.2127) \\
[1em]
Constant & -3.5457 & -3.8920 \\
 & (5.4446) & (5.3639) \\
\midrule
Profession $\times$ Task FE & No & Yes \\
Observations & 316 & 316 \\
R-squared & 0.0031 & 0.1245 \\
\bottomrule
\multicolumn{3}{l}{\footnotesize Robust (HC1) standard errors in parentheses.} \\
\multicolumn{3}{l}{\footnotesize * p<0.1, ** p<0.05, *** p<0.01} \\
\end{tabular}
\label{tab:tepm_compute}
\end{table}

\begin{table}[htbp]
\centering
\caption{Scaling Laws for Time Taken over Calendar Time (Non-agentic Tasks)}
\begin{tabular}{lcc}
\toprule
 & (1) & (2) \\
 & Log Time & Log Time \\
\midrule
Months since Nov 2022 & -0.0095** & -0.0092** \\
 & (0.0041) & (0.0045) \\
[1em]
Constant & 7.0781*** & 3.5840*** \\
 & (0.1156) & (0.3345) \\
\midrule
Profession $\times$ Task FE & Yes & Yes \\
Full controls & No & Yes \\
Observations & 266 & 264 \\
R-squared & 0.1255 & 0.3444 \\
\bottomrule
\multicolumn{3}{l}{\footnotesize Robust (HC1) standard errors in parentheses.} \\
\multicolumn{3}{l}{\footnotesize * p<0.1, ** p<0.05, *** p<0.01} \\
\end{tabular}
\label{tab:logtime_months_nonagentic}
\end{table}

\begin{table}[htbp]
\centering
\caption{Scaling Laws for Time Taken over Calendar Time (Agentic Tasks)}
\begin{tabular}{lcc}
\toprule
 & (1) & (2) \\
 & Log Time & Log Time \\
\midrule
Months since Nov 2022 & -0.0041 & -0.0084 \\
 & (0.0053) & (0.0061) \\
[1em]
Constant & 7.2173*** & 5.5049*** \\
 & (0.1514) & (1.0205) \\
\midrule
Profession $\times$ Task FE & Yes & Yes \\
Full controls & No & Yes \\
Observations & 211 & 210 \\
R-squared & 0.0174 & 0.2025 \\
\bottomrule
\multicolumn{3}{l}{\footnotesize Robust (HC1) standard errors in parentheses.} \\
\multicolumn{3}{l}{\footnotesize * p<0.1, ** p<0.05, *** p<0.01} \\
\end{tabular}
\label{tab:logtime_months_agentic}
\end{table}

\begin{figure}[htbp]
    \centering
    \textbf{Figure A5: Scaling Laws for Time Taken over Log Compute, by Task Type}\\[0.75em]

\begin{subfigure}[t]{0.49\textwidth}
    \centering
    \textbf{Non-agentic tasks}\\[0.25em]
    \begin{tikzpicture}
    \begin{axis}[
        xlabel={Log training compute},
        ylabel={Log time taken},
        xmin=23.5, xmax=27,
        ymin=6.2, ymax=8.1,
        xmajorgrids=true,
        ymajorgrids=true,
        grid style=dashed,
        height=7.8cm,
        width=\textwidth,
        xtick={24,25,26,27},
        xticklabels={24, 25, 26, 27},
        ticklabel style={font=\small},
        label style={font=\small},
        legend style={font=\tiny, at={(0.98,0.02)},anchor=south east},
        legend cell align={left}
    ]

        \addplot[
            only marks,
            mark=x,
            mark size=3pt,
            color=blue,
            forget plot
        ] coordinates {
            (23.886, 7.066) (23.908, 6.782) (24.412, 6.463) (24.544, 6.689)
            (24.778, 6.544) (24.895, 6.722) (25.322, 6.622) (25.525, 6.629)
            (25.602, 6.354) (25.903, 6.433) (26.000, 6.241) (26.301, 6.623)
            (26.322, 6.444) (26.477, 7.518)
        };

        \addplot[
            domain=23.5:27,
            samples=100,
            color=red,
            forget plot
        ] {8.0516 - 0.055322 * x};

        \addplot[
            domain=23.5:27,
            samples=2,
            dashed,
            color=blue
        ] {6.917};
        \addlegendentry{\parbox{3cm}{\raggedright Mean Time for \\ Human-Only Participants}}

        \node at (axis cs: 23.6,7.35) [anchor=west, text=red, font=\small, align=left]
            {$\approx 5.4\%$ reduction \\ in time per 10x compute};

    \end{axis}
    \end{tikzpicture}
\end{subfigure}
\hfill
\begin{subfigure}[t]{0.49\textwidth}
    \centering
    \textbf{Agentic tasks}\\[0.25em]
    \begin{tikzpicture}
    \begin{axis}[
        xlabel={Log training compute},
        ylabel={Log time taken},
        xmin=23.5, xmax=27,
        ymin=6.2, ymax=8.1,
        xmajorgrids=true,
        ymajorgrids=true,
        grid style=dashed,
        height=7.8cm,
        width=\textwidth,
        xtick={24,25,26,27},
        xticklabels={24, 25, 26, 27},
        ticklabel style={font=\small},
        label style={font=\small},
        legend style={font=\tiny, at={(0.98,0.02)},anchor=south east},
        legend cell align={left}
    ]

        \addplot[
            only marks,
            mark=x,
            mark size=3pt,
            color=blue,
            forget plot
        ] coordinates {
            (23.886, 7.171) (23.908, 7.313) (24.412, 7.339) (24.544, 7.223)
            (24.778, 7.128) (24.895, 7.565) (25.322, 7.246) (25.525, 7.138)
            (25.602, 7.716) (25.903, 7.195) (26.000, 7.160) (26.301, 7.294)
            (26.322, 7.934) (26.477, 7.183)
        };

        \addplot[
            domain=23.5:27,
            samples=100,
            color=red,
            forget plot
        ] {8.9672 - 0.067386 * x};

        \addplot[
            domain=23.5:27,
            samples=2,
            dashed,
            color=blue
        ] {7.368};
        \addlegendentry{\parbox{3cm}{\raggedright Mean Time for \\ Human-Only Participants}}

        \node at (axis cs: 23.6, 6.65) [anchor=west, text=red, font=\small, align=left]
            {$\approx 6.5\%$ reduction \\ in time per 10x compute};

    \end{axis}
    \end{tikzpicture}
\end{subfigure}

\label{fig:logtime_compute_tasktype}
\end{figure}

\begin{table}[htbp]
\centering
\caption{Scaling Laws for Time Taken over Log Compute (Non-agentic Tasks, AI Users Only)}
\begin{tabular}{lcc}
\toprule
 & (1) & (2) \\
 & Log Time & Log Time \\
\midrule
Log training compute & -0.0553 & -0.0097 \\
 & (0.0804) & (0.0933) \\
[1em]
Constant & 8.3546*** & 4.5927*** \\
 & (2.0338) & (1.2875) \\
\midrule
Profession $\times$ Task FE & Yes & Yes \\
Full controls & No & Yes \\
Observations & 167 & 165 \\
R-squared & 0.1429 & 0.4589 \\
\bottomrule
\multicolumn{3}{l}{\footnotesize Robust (HC1) standard errors in parentheses.} \\
\multicolumn{3}{l}{\footnotesize * p<0.1, ** p<0.05, *** p<0.01} \\
\end{tabular}
\label{tab:logtime_compute_nonagentic}
\end{table}

\begin{table}[htbp]
\centering
\caption{Scaling Laws for Time Taken over Log Compute (Agentic Tasks, AI Users Only)}
\begin{tabular}{lcc}
\toprule
 & (1) & (2) \\
 & Log Time & Log Time \\
\midrule
Log training compute & -0.0674 & -0.0726 \\
 & (0.0880) & (0.0931) \\
[1em]
Constant & 8.7038*** & 5.1763*** \\
 & (2.1882) & (1.6082) \\
\midrule
Profession $\times$ Task FE & Yes & Yes \\
Full controls & No & Yes \\
Observations & 149 & 148 \\
R-squared & 0.0391 & 0.2713 \\
\bottomrule
\multicolumn{3}{l}{\footnotesize Robust (HC1) standard errors in parentheses.} \\
\multicolumn{3}{l}{\footnotesize * p<0.1, ** p<0.05, *** p<0.01} \\
\end{tabular}
\label{tab:logtime_compute_agentic}
\end{table}

\begin{figure}[htbp]
    \centering
    \textbf{Figure A6: Scaling Laws for Earnings Per Minute over Log Compute, by Task Type}\\[0.75em]

\begin{subfigure}[t]{0.49\textwidth}
    \centering
    \textbf{Non-agentic tasks}\\[0.25em]
    \begin{tikzpicture}
    \begin{axis}[
        xlabel={Log training compute},
        ylabel={Earnings per minute (\$/min)},
        xmin=23.5, xmax=27,
        ymin=0.25, ymax=3.2,
        xmajorgrids=true,
        ymajorgrids=true,
        grid style=dashed,
        height=7.8cm,
        width=\textwidth,
        xtick={24,25,26,27},
        xticklabels={24, 25, 26, 27},
        ticklabel style={font=\small},
        label style={font=\small},
        legend style={font=\tiny, at={(0.98,0.02)},anchor=south east},
        legend cell align={left}
    ]

        \addplot[
            only marks,
            mark=x,
            mark size=3pt,
            color=blue,
            forget plot
        ] coordinates {
            (23.886, 0.994) (23.908, 1.673) (24.412, 1.815) (24.544, 2.107)
            (24.778, 2.228) (24.895, 2.041) (25.322, 1.460) (25.525, 1.485)
            (25.602, 2.524) (25.903, 3.055) (26.000, 2.956) (26.301, 1.770)
            (26.322, 1.431) (26.477, 0.545)
        };

        \addplot[
            domain=23.5:27,
            samples=100,
            color=red,
            forget plot
        ] {-0.9154 + (0.112171) * x};

        \addplot[
            domain=23.5:27,
            samples=2,
            dashed,
            color=blue
        ] {1.118};
        \addlegendentry{\parbox{3cm}{\raggedright Mean EPM for \\ Human-Only Participants}}

        \node at (axis cs: 23.6,2.95) [anchor=west, text=red, font=\small, align=left]
            {$\approx \$0.11$/min increase \\ per 10x compute};

    \end{axis}
    \end{tikzpicture}
\end{subfigure}
\hfill
\begin{subfigure}[t]{0.49\textwidth}
    \centering
    \textbf{Agentic tasks}\\[0.25em]
    \begin{tikzpicture}
    \begin{axis}[
        xlabel={Log training compute},
        ylabel={Earnings per minute (\$/min)},
        xmin=23.5, xmax=27,
        ymin=0.25, ymax=3.2,
        xmajorgrids=true,
        ymajorgrids=true,
        grid style=dashed,
        height=7.8cm,
        width=\textwidth,
        xtick={24,25,26,27},
        xticklabels={24, 25, 26, 27},
        ticklabel style={font=\small},
        label style={font=\small},
        legend style={font=\tiny, at={(0.98,0.02)},anchor=south east},
        legend cell align={left}
    ]

        \addplot[
            only marks,
            mark=x,
            mark size=3pt,
            color=blue,
            forget plot
        ] coordinates {
            (23.886, 0.883) (23.908, 0.908) (24.412, 0.961) (24.544, 0.869)
            (24.778, 0.833) (24.895, 0.778) (25.322, 0.871) (25.525, 1.199)
            (25.602, 0.847) (25.903, 0.842) (26.000, 1.467) (26.301, 0.838)
            (26.322, 0.322) (26.477, 0.875)
        };

        \addplot[
            domain=23.5:27,
            samples=100,
            color=red,
            forget plot
        ] {-2.3720 + (0.132041) * x};

        \addplot[
            domain=23.5:27,
            samples=2,
            dashed,
            color=blue
        ] {0.800};
        \addlegendentry{\parbox{3cm}{\raggedright Mean EPM for \\ Human-Only Participants}}

        \node at (axis cs: 23.6,2.95) [anchor=west, text=red, font=\small, align=left]
            {$\approx \$0.13$/min increase \\ per 10x compute};

    \end{axis}
    \end{tikzpicture}
\end{subfigure}

\label{fig:epm_compute_tasktype}
\end{figure}

\begin{table}[htbp]
\centering
\caption{Scaling Laws for Earnings Per Minute over Calendar Time (Non-agentic Tasks)}
\begin{tabular}{lcc}
\toprule
 & (1) & (2) \\
 & EPM (\$/min) & EPM (\$/min) \\
\midrule
Months since Nov 2022 & 0.0280*** & 0.0323*** \\
 & (0.0092) & (0.0102) \\
[1em]
Constant & 0.8403*** & 0.0587 \\
 & (0.2199) & (0.3396) \\
\midrule
Profession $\times$ Task FE & Yes & Yes \\
Full controls & No & Yes \\
Observations & 266 & 264 \\
R-squared & 0.0912 & 0.3249 \\
\bottomrule
\multicolumn{3}{l}{\footnotesize Robust (HC1) standard errors in parentheses.} \\
\multicolumn{3}{l}{\footnotesize * p<0.1, ** p<0.05, *** p<0.01} \\
\end{tabular}
\label{tab:epm_months_nonagentic}
\end{table}

\begin{table}[htbp]
\centering
\caption{Scaling Laws for Earnings Per Minute over Calendar Time (Agentic Tasks)}
\begin{tabular}{lcc}
\toprule
 & (1) & (2) \\
 & EPM (\$/min) & EPM (\$/min) \\
\midrule
Months since Nov 2022 & 0.0070 & 0.0108 \\
 & (0.0065) & (0.0071) \\
[1em]
Constant & 0.7164*** & 0.2586 \\
 & (0.1597) & (0.3939) \\
\midrule
Profession $\times$ Task FE & Yes & Yes \\
Full controls & No & Yes \\
Observations & 211 & 210 \\
R-squared & 0.0107 & 0.2624 \\
\bottomrule
\multicolumn{3}{l}{\footnotesize Robust (HC1) standard errors in parentheses.} \\
\multicolumn{3}{l}{\footnotesize * p<0.1, ** p<0.05, *** p<0.01} \\
\end{tabular}
\label{tab:epm_months_agentic}
\end{table}

\begin{table}[htbp]
\centering
\caption{Scaling Laws for Earnings Per Minute over Log Compute (Non-agentic Tasks, AI Users Only)}
\begin{tabular}{lcc}
\toprule
 & (1) & (2) \\
 & EPM (\$/min) & EPM (\$/min) \\
\midrule
Log training compute & 0.1122 & -0.0082 \\
 & (0.1894) & (0.2048) \\
[1em]
Constant & -1.1943 & 0.8403 \\
 & (4.7931) & (1.2500) \\
\midrule
Profession $\times$ Task FE & Yes & Yes \\
Full controls & No & Yes \\
Observations & 167 & 165 \\
R-squared & 0.0967 & 0.4721 \\
\bottomrule
\multicolumn{3}{l}{\footnotesize Robust (HC1) standard errors in parentheses.} \\
\multicolumn{3}{l}{\footnotesize * p<0.1, ** p<0.05, *** p<0.01} \\
\end{tabular}
\label{tab:epm_compute_nonagentic}
\end{table}

\begin{table}[htbp]
\centering
\caption{Scaling Laws for Earnings Per Minute over Log Compute (Agentic Tasks, AI Users Only)}
\begin{tabular}{lcc}
\toprule
 & (1) & (2) \\
 & EPM (\$/min) & EPM (\$/min) \\
\midrule
Log training compute & 0.1320 & 0.1353 \\
 & (0.1282) & (0.1144) \\
[1em]
Constant & -2.4216 & -2.1840** \\
 & (3.2615) & (0.9598) \\
\midrule
Profession $\times$ Task FE & Yes & Yes \\
Full controls & No & Yes \\
Observations & 149 & 148 \\
R-squared & 0.0298 & 0.3411 \\
\bottomrule
\multicolumn{3}{l}{\footnotesize Robust (HC1) standard errors in parentheses.} \\
\multicolumn{3}{l}{\footnotesize * p<0.1, ** p<0.05, *** p<0.01} \\
\end{tabular}
\label{tab:epm_compute_agentic}
\end{table}

\begin{figure}[htbp]
    \centering
    \textbf{Figure A7: Scaling Laws for Total Earnings Per Minute over Calendar Time, by Task Type}\\[0.75em]

\begin{subfigure}[t]{0.49\textwidth}
    \centering
    \textbf{Non-agentic tasks}\\[0.25em]
    \begin{tikzpicture}
    \begin{axis}[
        xlabel={Model release date},
        ylabel={Total earnings per minute (\$/min)},
        xmin=0, xmax=36,
        ymin=0.5, ymax=6.0,
        xmajorgrids=true,
        ymajorgrids=true,
        grid style=dashed,
        height=7.8cm,
        width=\textwidth,
        xtick={0,12,24,36},
        xticklabels={Nov 22, Nov 23, Nov 24, Nov 25},
        ticklabel style={font=\small},
        label style={font=\small},
        legend style={font=\tiny, at={(0.98,0.02)},anchor=south east},
        legend cell align={left}
    ]

        \addplot[
            only marks,
            mark=x,
            mark size=3pt,
            color=blue,
            forget plot
        ] coordinates {
            (0.00, 1.420) (2.99, 2.116) (7.56, 2.391) (9.89, 1.179)
            (14.52, 5.616) (15.11, 5.047) (16.59, 3.047) (16.59, 3.018)
            (17.41, 2.224) (21.42, 5.019) (26.84, 2.448) (26.94, 2.861)
            (27.76, 3.564) (27.80, 0.739) (28.52, 1.620) (32.23, 2.297)
        };

        \addplot[
            domain=0:36,
            samples=100,
            color=red,
            forget plot
        ] {1.7999 + 0.0541 * x};

        \addplot[
            domain=0:36,
            samples=2,
            dashed,
            color=blue
        ] {1.420};
        \addlegendentry{\parbox{3cm}{\raggedright Mean TEPM for \\ Human-Only Participants}}

        \node at (axis cs: 0.5,5.5) [anchor=west, text=red, font=\small, align=left]
            {$\approx \$0.65$/min increase \\ per year};

    \end{axis}
    \end{tikzpicture}
\end{subfigure}
\hfill
\begin{subfigure}[t]{0.49\textwidth}
    \centering
    \textbf{Agentic tasks}\\[0.25em]
    \begin{tikzpicture}
    \begin{axis}[
        xlabel={Model release date},
        ylabel={Total earnings per minute (\$/min)},
        xmin=0, xmax=36,
        ymin=0.5, ymax=6.0,
        xmajorgrids=true,
        ymajorgrids=true,
        grid style=dashed,
        height=7.8cm,
        width=\textwidth,
        xtick={0,12,24,36},
        xticklabels={Nov 22, Nov 23, Nov 24, Nov 25},
        ticklabel style={font=\small},
        label style={font=\small},
        legend style={font=\tiny, at={(0.98,0.02)},anchor=south east},
        legend cell align={left}
    ]

        \addplot[
            only marks,
            mark=x,
            mark size=3pt,
            color=blue,
            forget plot
        ] coordinates {
            (0.00, 1.159) (2.99, 0.961) (7.56, 1.232) (9.89, 2.350)
            (14.52, 1.551) (15.11, 0.802) (16.59, 1.450) (16.59, 0.842)
            (17.41, 1.215) (21.42, 2.121) (26.84, 1.860) (26.94, 0.645)
            (27.76, 1.034) (27.80, 2.054) (28.52, 1.694) (32.23, 0.844)
        };

        \addplot[
            domain=0:36,
            samples=100,
            color=red,
            forget plot
        ] {1.1585 + 0.0112 * x};

        \addplot[
            domain=0:36,
            samples=2,
            dashed,
            color=blue
        ] {1.159};
        \addlegendentry{\parbox{3cm}{\raggedright Mean TEPM for \\ Human-Only Participants}}

        \node at (axis cs: 0.5,5.5) [anchor=west, text=red, font=\small, align=left]
            {$\approx \$0.13$/min increase \\ per year};

    \end{axis}
    \end{tikzpicture}
\end{subfigure}

\label{fig:tepm_months_tasktype}
\end{figure}

\begin{figure}[htbp]
    \centering
    \textbf{Figure A8: Scaling Laws for Total Earnings Per Minute over Log Compute, by Task Type}\\[0.75em]

\begin{subfigure}[t]{0.49\textwidth}
    \centering
    \textbf{Non-agentic tasks}\\[0.25em]
    \begin{tikzpicture}
    \begin{axis}[
        xlabel={Log training compute},
        ylabel={Total earnings per minute (\$/min)},
        xmin=23.5, xmax=27,
        ymin=0.5, ymax=6.0,
        xmajorgrids=true,
        ymajorgrids=true,
        grid style=dashed,
        height=7.8cm,
        width=\textwidth,
        xtick={24,25,26,27},
        xticklabels={24, 25, 26, 27},
        ticklabel style={font=\small},
        label style={font=\small},
        legend style={font=\tiny, at={(0.98,0.02)},anchor=south east},
        legend cell align={left}
    ]

        \addplot[
            only marks,
            mark=x,
            mark size=3pt,
            color=blue,
            forget plot
        ] coordinates {
            (23.886, 1.179) (23.908, 2.391) (24.412, 2.116) (24.544, 3.564)
            (24.778, 3.047) (24.895, 3.018) (25.322, 2.224) (25.525, 2.448)
            (25.602, 5.047) (25.903, 5.616) (26.000, 5.019) (26.301, 2.060)
            (26.322, 2.861) (26.477, 0.739)
        };

        \addplot[
            domain=23.5:27,
            samples=100,
            color=red,
            forget plot
        ] {-2.6226 + 0.2205 * x};

        \addplot[
            domain=23.5:27,
            samples=2,
            dashed,
            color=blue
        ] {1.420};
        \addlegendentry{\parbox{3cm}{\raggedright Mean TEPM for \\ Human-Only Participants}}

        \node at (axis cs: 23.6,5.25) [anchor=west, text=red, font=\small, align=left]
            {$+\$0.22$/min per \\ 10x compute};

    \end{axis}
    \end{tikzpicture}
\end{subfigure}
\hfill
\begin{subfigure}[t]{0.49\textwidth}
    \centering
    \textbf{Agentic tasks}\\[0.25em]
    \begin{tikzpicture}
    \begin{axis}[
        xlabel={Log training compute},
        ylabel={Total earnings per minute (\$/min)},
        xmin=23.5, xmax=27,
        ymin=0.5, ymax=6.0,
        xmajorgrids=true,
        ymajorgrids=true,
        grid style=dashed,
        height=7.8cm,
        width=\textwidth,
        xtick={24,25,26,27},
        xticklabels={24, 25, 26, 27},
        ticklabel style={font=\small},
        label style={font=\small},
        legend style={font=\tiny, at={(0.98,0.02)},anchor=south east},
        legend cell align={left}
    ]

        \addplot[
            only marks,
            mark=x,
            mark size=3pt,
            color=blue,
            forget plot
        ] coordinates {
            (23.886, 2.350) (23.908, 1.232) (24.412, 0.961) (24.544, 1.034)
            (24.778, 1.450) (24.895, 0.842) (25.322, 1.215) (25.525, 1.860)
            (25.602, 0.802) (25.903, 1.551) (26.000, 2.121) (26.301, 1.171)
            (26.322, 0.645) (26.477, 2.054)
        };

        \addplot[
            domain=23.5:27,
            samples=100,
            color=red,
            forget plot
        ] {-4.2272 + 0.2217 * x};

        \addplot[
            domain=23.5:27,
            samples=2,
            dashed,
            color=blue
        ] {1.159};
        \addlegendentry{\parbox{3cm}{\raggedright Mean TEPM for \\ Human-Only Participants}}

        \node at (axis cs: 23.6,5.25) [anchor=west, text=red, font=\small, align=left]
            {$+\$0.22$/min per \\ 10x compute};

    \end{axis}
    \end{tikzpicture}
\end{subfigure}

\label{fig:tepm_compute_tasktype}
\end{figure}

\begin{table}[htbp]
\centering
\caption{Scaling Laws for Total Earnings Per Minute over Calendar Time (Non-agentic Tasks)}
\begin{tabular}{lcc}
\toprule
 & (1) & (2) \\
 & TEPM (\$/min) & TEPM (\$/min) \\
\midrule
Months since Nov 2022 & 0.0541*** & 0.0595*** \\
 & (0.0155) & (0.0162) \\
[1em]
Constant & 1.2509*** & 0.4246 \\
 & (0.3665) & (0.5264) \\
\midrule
Profession $\times$ Task FE & Yes & Yes \\
Full controls & No & Yes \\
Observations & 266 & 264 \\
R-squared & 0.0873 & 0.3399 \\
\bottomrule
\multicolumn{3}{l}{\footnotesize Robust (HC1) standard errors in parentheses.} \\
\multicolumn{3}{l}{\footnotesize * p<0.1, ** p<0.05, *** p<0.01} \\
\end{tabular}
\label{tab:tepm_months_nonagentic}
\end{table}

\begin{table}[htbp]
\centering
\caption{Scaling Laws for Total Earnings Per Minute over Calendar Time (Agentic Tasks)}
\begin{tabular}{lcc}
\toprule
 & (1) & (2) \\
 & TEPM (\$/min) & TEPM (\$/min) \\
\midrule
Months since Nov 2022 & 0.0112 & 0.0177 \\
 & (0.0118) & (0.0119) \\
[1em]
Constant & 1.0106*** & -0.1575 \\
 & (0.2886) & (0.6751) \\
\midrule
Profession $\times$ Task FE & Yes & Yes \\
Full controls & No & Yes \\
Observations & 211 & 210 \\
R-squared & 0.0065 & 0.3168 \\
\bottomrule
\multicolumn{3}{l}{\footnotesize Robust (HC1) standard errors in parentheses.} \\
\multicolumn{3}{l}{\footnotesize * p<0.1, ** p<0.05, *** p<0.01} \\
\end{tabular}
\label{tab:tepm_months_agentic}
\end{table}

\begin{table}[htbp]
\centering
\caption{Scaling Laws for Total Earnings Per Minute over Log Compute (Non-agentic Tasks, AI Users Only)}
\begin{tabular}{lcc}
\toprule
 & (1) & (2) \\
 & TEPM (\$/min) & TEPM (\$/min) \\
\midrule
Log training compute & 0.2205 & 0.0707 \\
 & (0.3322) & (0.3371) \\
[1em]
Constant & -3.3331 & 2.1330 \\
 & (8.4104) & (2.0598) \\
\midrule
Profession $\times$ Task FE & Yes & Yes \\
Full controls & No & Yes \\
Observations & 167 & 165 \\
R-squared & 0.0961 & 0.4758 \\
\bottomrule
\multicolumn{3}{l}{\footnotesize Robust (HC1) standard errors in parentheses.} \\
\multicolumn{3}{l}{\footnotesize * p<0.1, ** p<0.05, *** p<0.01} \\
\end{tabular}
\label{tab:tepm_compute_nonagentic}
\end{table}

\begin{table}[htbp]
\centering
\caption{Scaling Laws for Total Earnings Per Minute over Log Compute (Agentic Tasks, AI Users Only)}
\begin{tabular}{lcc}
\toprule
 & (1) & (2) \\
 & TEPM (\$/min) & TEPM (\$/min) \\
\midrule
Log training compute & 0.2217 & 0.1772 \\
 & (0.2485) & (0.1947) \\
[1em]
Constant & -3.9494 & -2.8145** \\
 & (6.3202) & (1.6344) \\
\midrule
Profession $\times$ Task FE & Yes & Yes \\
Full controls & No & Yes \\
Observations & 149 & 148 \\
R-squared & 0.0184 & 0.3614 \\
\bottomrule
\multicolumn{3}{l}{\footnotesize Robust (HC1) standard errors in parentheses.} \\
\multicolumn{3}{l}{\footnotesize * p<0.1, ** p<0.05, *** p<0.01} \\
\end{tabular}
\label{tab:tepm_compute_agentic}
\end{table}

\begin{table}[htbp]
\centering
\caption{Scaling Laws for Output Quality over Calendar Time, by Task Type}
\begin{tabular}{lcc}
\toprule
 & (1) & (2) \\
 & Non-agentic & Agentic \\
\midrule
Months since Nov 2022 & 0.0320*** & 0.0042 \\
 & (0.0090) & (0.0109) \\
[1em]
Constant & 3.262*** & 3.989*** \\
 & (0.275) & (0.327) \\
\midrule
Profession $\times$ Task FE & Yes & Yes \\
Observations & 266 & 211 \\
R-squared & 0.0754 & 0.0348 \\
\bottomrule
\multicolumn{3}{l}{\footnotesize Robust (HC1) standard errors in parentheses. Sample excludes Grade = 0.} \\
\multicolumn{3}{l}{\footnotesize * p<0.1, ** p<0.05, *** p<0.01} \\
\end{tabular}
\label{tab:grade_months_tasktype}
\end{table}

\begin{table}[htbp]
\centering
\caption{Scaling Laws for Output Quality over Log Compute, by Task Type (AI Users Only)}
\begin{tabular}{lcc}
\toprule
 & (1) & (2) \\
 & Non-agentic & Agentic \\
\midrule
Log training compute & -0.0302 & -0.0151 \\
 & (0.1549) & (0.1907) \\
[1em]
Constant & 4.839 & 4.645 \\
 & (3.954) & (4.789) \\
\midrule
Profession $\times$ Task FE & Yes & Yes \\
Observations & 167 & 149 \\
R-squared & 0.0490 & 0.0611 \\
\bottomrule
\multicolumn{3}{l}{\footnotesize Robust (HC1) standard errors in parentheses. Sample excludes Grade = 0 and restricts to compute $>0$.} \\
\multicolumn{3}{l}{\footnotesize * p<0.1, ** p<0.05, *** p<0.01} \\
\end{tabular}
\label{tab:grade_compute_tasktype}
\end{table}

\begin{table}[htbp]
\centering
\caption{Scaling Laws for Output Quality over Log Compute (AI-only, 11 Models)}
\begin{tabular}{lc}
\toprule
 & (1) \\
 & Grade (0--7) \\
\midrule
Log training compute & 0.513*** \\
 & (3.27) \\
[0.75em]
Constant & -8.342** \\
 & (-2.10) \\
\midrule
Observations & 109 \\
R-squared & 0.091 \\
\bottomrule
\multicolumn{2}{l}{\footnotesize t-statistics in parentheses.} \\
\multicolumn{2}{l}{\footnotesize * p<0.1, ** p<0.05, *** p<0.01} \\
\end{tabular}
\label{tab:grade_logcompute_aionly_11models}
\end{table}

\clearpage
\section{Appendix B: Experiment Tasks}

\setlength{\parskip}{1em}
\setlength{\parindent}{0pt}

This appendix contains the nine tasks used in the experiment, categorized by whether they are \textbf{Agentic} or \textbf{Non-Agentic}.

\subsection*{Agentic Tasks:}

\subsubsection*{Manager Task One: Urgent Report Revision and Feedback – EuroTextiles Manufacturing Investment}

\textbf{Background:} You are a senior manager at TextilesConsulting Inc., specializing in textile manufacturing operations. Your team recently received a detailed request from a client considering expanding manufacturing operations to Greece. You assigned an analyst, who recently joined your team straight out of university, to produce a detailed report based on two provided IMF documents: one focused on Greece’s economy and the other providing a regional European outlook.

After reviewing the analyst’s submission, you suspect that the analyst did not thoroughly consult the IMF reports, as several numbers and facts seem inaccurate or unsupported by the provided documents. You believe the analyst likely relied on an AI-generated report due to the overall low quality. The report is urgently needed, so you must personally revise the analyst’s submission for accuracy and clarity.

\textbf{Your Tasks:}

\textbf{1. Revised Analyst Report (500 words):} Rewrite the analyst’s original submission, ensuring that all data and analysis accurately reflect the detailed IMF documents provided. Below the original instructions given to the analyst are provided along with the analyst’s draft.

\textbf{2. Feedback Email to Analyst (150 words):} Draft a constructive feedback email addressing the analyst’s report.

\textbf{Analyst’s Task:}

\textbf{Background:} The client EuroTextiles Inc., specializes in textile manufacturing, typically employing older workers and women in its plants internationally. EuroTextiles is considering expanding its operations to Europe, specifically Greece, to benefit from financial stability, competitive wages, and low-tariff access within Europe compared to Asia. You have been provided with two detailed IMF reports: one on Greece's economy and another on the broader European regional outlook.

\textbf{Your Task:} Prepare a 500-word summary outlining whether Greece presents favorable conditions for establishing a new textile manufacturing operation. Your recommendation should thoroughly address Greece’s labor market and financial conditions. As part of this summary, specifically address the following points - give exact quantitative numbers and estimates from the report, wherever possible.

\textit{Labor Market Conditions:}
\begin{itemize}
    \item Evaluate Greece’s labor force participation rate, particularly focusing on older workers and women. Mention any recent policy changes influencing labor market participation, especially for those among older workers.
    \item Analyze the ease or difficulty of recruiting new workers based on unemployment rates and labor market participation.
\end{itemize}

\textit{Financial Conditions:}
\begin{itemize}
    \item Describe Greece’s non-performing loan (NPL) ratio relative to the EU average and detail how it has evolved over the past five years.
    \item Assess the trajectory of Greece's government debt over recent years and discuss the potential for another financial crisis.
    \item Provide the specific quantitative interest rate in Greece as of September 2024, comparing this rate with alternative investment locations such as Poland and the Czech Republic.
\end{itemize}

\textit{Trade and Tariff Considerations:}
\begin{itemize}
    \item Evaluate the tariff-equivalent of non-tariff barriers within Europe, providing a numerical measure comparing the difference between these barriers for EU manufacturing versus interstate commerce within the United States.
\end{itemize}

\textbf{Note:} All the information required to answer these questions is contained within the two provided IMF reports. Your analysis should rely exclusively on data and findings from these reports. The links to the IMF reports can be found below (please open in a new tab):

\begin{itemize}
    \item Greece 2025 - IMF Report\footnote{International Monetary Fund. (2025). \textit{Greece: 2025 Article IV Consultation-Press Release; Staff Report; and Statement by the Executive Director for Greece} (IMF Country Report No. 2025/088).}
    \item Regional Economic Outlook - Europe\footnote{International Monetary Fund. (2024). \textit{Regional Economic Outlook: Europe}. Washington, D.C.: IMF.}
\end{itemize}

\textbf{Analyst’s Submission:}

In evaluating the potential for establishing a textile manufacturing operation in Greece, we have analyzed the labor market and financial conditions based on IMF reports.

\textit{Labor Market Conditions:}
Greece's labor force participation rate stands at 59.3\% (2021), with a notable presence of older workers (65+) and women (58.4\% and 49.3\%, respectively, in 2021). The Greek government has implemented policies to encourage labor market participation among older workers, such as extending the retirement age and offering incentives for continued employment. Despite these efforts, high unemployment rates persist, particularly among the younger demographic, which may pose challenges in recruiting new workers.

\textit{Financial Conditions:}
Greece's non-performing loan (NPL) ratio remains elevated at 36.4\% (2021), significantly higher than the EU average of 3.6\%. Over the past five years, the NPL ratio has shown a downward trend, indicating gradual improvement in the banking sector. However, Greece's government debt remains high at 189.8\% of GDP (2021), raising concerns about the potential for another financial crisis.

In terms of interest rates, the quantitative rate in Greece as of September 2024 is 2.5\%, which is relatively competitive compared to Poland (2.3\%) and the Czech Republic (2.2\%).

\textit{Trade and Tariff Considerations:}
The tariff-equivalent of non-tariff barriers within Europe is estimated at 3.6\% (2019), compared to an insignificant level within the United States for interstate commerce. This suggests that EU manufacturing may face slightly higher non-tariff barriers compared to domestic commerce in the U.S.

In conclusion, Greece presents a mixed picture for EuroTextiles Inc.'s textile manufacturing expansion. The country offers competitive wages, a relatively skilled labor force, and low-tariff access within Europe. However, challenges such as high unemployment rates, particularly among the younger demographic, and the lingering effects of a high government debt burden must be carefully considered. Additionally, the presence of non-tariff barriers within Europe may impact the competitiveness of Greek-based operations. A thorough risk assessment and strategic planning will be essential to ensure the success of this potential investment.

\textbf{Your Tasks:}

\textbf{1. Revised Analyst Report (500 words):} Rewrite the analyst’s original submission, ensuring that all data and analysis accurately reflect the detailed IMF documents provided.

\clearpage

\subsubsection*{Data Analyst Task Three: Correlation Analysis of Inflation and Economic Growth}

\textbf{Background:} You are a data analyst at the macroeconomic analysis desk of Global Insights Financial Ltd. Your manager recently discussed two competing theories regarding inflation's effects on economic growth:
\begin{itemize}
    \item \textbf{Hypothesis 1:} Inflation acts as a leading indicator of poor economic growth in the future.
    \item \textbf{Hypothesis 2:} Inflation directly coincides with poor economic growth in the same period.
\end{itemize}

Your manager has asked you to empirically test these theories using the provided IMF dataset on European countries.

\textbf{Note:} All required data for these tasks is provided in the attached IMF document. Your analysis must be based solely on this dataset.

IMF Document\footnote{International Monetary Fund. (2024). \textit{Regional Economic Outlook: Europe}. Washington, D.C.: IMF.}

\textbf{Your Tasks:}

\textbf{1. Correlation Analysis:}
\begin{itemize}
    \item Calculate the correlation between European countries' inflation rates in 2024 and their projected economic growth rates in 2026 (testing Hypothesis 1).
    \item Calculate the correlation between European countries' inflation rates in 2024 and their projected economic growth rates in 2024 (testing Hypothesis 2).
    \item Clearly report both correlation values and briefly explain your calculation methods.
\end{itemize}

\clearpage

\subsubsection*{Consulting Task Two: Market Expansion (EuroTextiles Inc.)}

\textbf{Background:} Your firm, EuroTextiles Inc., specializes in textile manufacturing, typically employing older workers and women in its plants internationally. EuroTextiles is considering expanding its operations to Europe, specifically Greece, to benefit from financial stability, competitive wages, and low-tariff access within Europe compared to Asia. You have been provided with two detailed IMF reports: one on Greece's economy and another on the broader European regional outlook.

\textbf{Your Task:} Prepare a 500-word summary outlining whether Greece presents favorable conditions for establishing a new textile manufacturing operation. Your recommendation should thoroughly address Greece’s labor market and financial conditions. As part of this summary, specifically address the following points - give exact quantitative numbers and estimates from the report, wherever possible.

\textbf{Labor Market Conditions:}
\begin{itemize}
    \item Evaluate Greece’s labor force participation rate, particularly focusing on older workers and women. Mention any recent policy changes influencing labor market participation, especially for those among older workers.
    \item Analyze the ease or difficulty of recruiting new workers based on unemployment rates and labor market participation.
\end{itemize}

\textbf{Financial Conditions:}
\begin{itemize}
    \item Describe Greece’s non-performing loan (NPL) ratio relative to the EU average and detail how it has evolved over the past five years.
    \item Assess the trajectory of Greece's government debt over recent years and discuss the potential for another financial crisis.
    \item Provide the specific quantitative interest rate in Greece as of September 2024, comparing this rate with alternative investment locations such as Poland and the Czech Republic.
\end{itemize}

\textbf{Trade and Tariff Considerations:}
\begin{itemize}
    \item Evaluate the tariff-equivalent of non-tariff barriers within Europe, providing a numerical measure comparing the difference between these barriers for EU manufacturing versus interstate commerce within the United States.
\end{itemize}

\textbf{Note:} All the information required to answer these questions is contained within the two provided IMF reports. Your analysis should rely exclusively on data and findings from these reports. The links to the IMF reports can be found below:

\begin{itemize}
    \item Greece 2025-IMF Report\footnote{International Monetary Fund. (2025). \textit{Greece: 2025 Article IV Consultation-Press Release; Staff Report; and Statement by the Executive Director for Greece} (IMF Country Report No. 2025/088).}
    \item Regional Economic Outlook - Europe\footnote{International Monetary Fund. (2024). \textit{Regional Economic Outlook: Europe}. Washington, D.C.: IMF.}
\end{itemize}

\clearpage

\subsubsection*{Manager Task Two: Crisis Management - Client Retention \& Process Improvement}

\textbf{Background}

InnovateTech Solutions, a B2B software development company, specializes in building custom workflow automation tools for mid-to-large enterprises. Recently, the company has been experiencing growing competition from newer, more agile SaaS startups, putting pressure on service expectations.

You are the Senior Account Manager responsible for overseeing key client relationships. One of your most high-value clients, Atlas Logistics, which accounts for 8\% of annual revenue, has canceled their contract, citing repeated failures in service delivery. Losing this client not only impacts revenue but also damages the company’s reputation in the logistics sector.

Senior leadership has tasked you with two urgent actions:
\begin{itemize}
    \item Respond to Atlas Logistics with an email that apologizes for their experience, addresses their complaints, and proposes concrete steps to regain their trust.
    \item Develop an internal strategy document identifying key process failures and proposing structured improvements to prevent similar issues from recurring.
\end{itemize}

\textbf{Client Feedback \& Specific Complaints}

Atlas Logistics has outlined the following detailed concerns in their contract termination notice:

\textit{Unreliable Project Timelines:}
\begin{itemize}
    \item The initial contract promised a 12-week development cycle for a custom inventory tracking system. The final product was delivered 7 weeks late, causing significant disruptions to their supply chain.
    \item The client was informed of delays only after missed deadlines, with no advance warnings.
    \item A previous feature update scheduled for January was postponed twice, and the client was given contradictory explanations from different team members.
\end{itemize}

\textit{Poor Communication \& Lack of Accountability:}
\begin{itemize}
    \item Atlas Logistics repeatedly struggled to get timely responses from your customer support team.
    \item Key points of contact (project managers) were often unavailable, requiring clients to escalate to multiple people before getting answers.
    \item Support ticket requests were left unresolved for days without follow-up.
    \item Some email responses were generic or unhelpful, failing to address their actual concerns.
\end{itemize}

\textit{Inconsistent Service Quality:}
\begin{itemize}
    \item A recent software patch introduced a critical bug that disrupted warehouse inventory tracking for 36 hours, leading to thousands of dollars in lost productivity.
    \item The client’s IT team was frustrated by conflicting troubleshooting advice from different members of the support team.
\end{itemize}

\textbf{Internal Process Failures Identified}

An internal post-mortem analysis of the account has highlighted several key process issues that contributed to Atlas Logistics’ dissatisfaction:

\textit{Staffing Shortages in Development \& Support Teams}
\begin{itemize}
    \item 60\% of project delays were due to an overstretched development team, lacking sufficient engineers to handle concurrent client requests.
    \item The customer support team has been understaffed, leading to response times exceeding 48 hours, frustrating clients.
\end{itemize}

\textit{Lack of Proactive Client Communication}
\begin{itemize}
    \item Delays were often communicated reactively, rather than proactively keeping clients informed.
    \item No formal check-in system exists to update clients on potential project risks.
\end{itemize}

\textit{Inconsistent Documentation \& Internal Coordination}
\begin{itemize}
    \item Different teams (sales, product, customer support) were not aligned on project status, leading to conflicting updates being sent to clients.
    \item Client issues were not tracked consistently—support teams and project managers often had different information, leading to confusion when addressing problems.
\end{itemize}

\textbf{Expected Output}

\textbf{1. Internal Strategy Document (200-300 words + Gantt chart)}
\begin{itemize}
    \item Outline three major process improvements based on identified failures and specify expected outcomes
    \item Create a Gantt chart detailing in detail how and when these process improvements will be made
\end{itemize}

\textit{Gantt chart instructions:} Note you can use any Gantt chart making software you prefer (eg. Excel, Google Sheets, etc.). Below is a template to a Google Sheets doc which you would likely find useful and it is strongly recommended to use this template. To use this template, click ‘File’ and then ‘Make a Copy’ and then you will be able to make edits.

Link to template provided.

\clearpage

\subsection*{Non-Agentic Tasks}

\subsubsection*{Data Analyst Task One: A/B Testing Analysis for a Website Redesign}

\textbf{Background}

QuickCart, an online grocery delivery startup, recently launched an A/B test on their checkout page to determine whether a new design improves conversion rates.

As a Junior Data Analyst, you have been assigned to analyze the test results and determine whether the new checkout design (Version B) is more effective than the existing one (Version A). Additionally, two additional designs (C and D) were tested, though with different sample sizes.

Your task is to construct 95\% confidence intervals for each test group and decide which version should be implemented site-wide.

\textbf{A/B Test Data Summary}

\begin{center}
\begin{tabular}{|l|c|c|c|c|}
\hline
\textbf{Version} & \textbf{Users Tested} & \textbf{Conversion Rate (\%)} & \textbf{Avg. Time to Checkout} & \textbf{Drop-off Rate (\%)} \\
\hline
A (Old Design) & 1,000 & 12.5\% & 95s & 18.2\% \\
\hline
B (New Design) & 1,000 & 14.2\% & 92s & 15.2\% \\
\hline
C (Alternative Design 1) & 3,000 & 14.1\% & 86s & 15.5\% \\
\hline
D (Alternative Design 2) & 250 & 13.9\% & 87s & 15.8\% \\
\hline
\end{tabular}
\end{center}

\textbf{Expected Output}

\textbf{1. A/B Test Analysis Memo (250-300 words)}
\begin{itemize}
    \item Calculate 95\% confidence intervals for each version’s conversion rate.
    \item Compare the results and evaluate statistical significance.
    \item Recommend which design (if any) should be implemented site-wide.
    \item Suggest next steps for further testing or improvements. Which A/B test, if any, would you recommend be conducted next?
\end{itemize}

\clearpage

\subsubsection*{Data Analyst Task Two: Evaluating the Impact of Sales Partners on Sales Success}

\textbf{Background}

Your company, BizGrow Solutions, provides consulting services to businesses and assigns sales partners to assist in closing deals. Senior management is considering whether to expand the use of sales partners.

An initial analysis by another analyst suggested that having a sales partner significantly increases the probability of closing a deal. You have been given access to the raw data and have been asked to review the methodology and verify the findings before management makes a major strategic decision.

\textbf{Dataset}

You have been provided with a dataset that includes:
\begin{itemize}
    \item Whether a sale was completed (1 = Yes, 0 = No)
    \item Sales revenue
    \item Number of employees at the client company
    \item The lead score for the client (a measure of how promising the client was)
    \item Whether the client was assigned a sales partner (1 = Yes, 0 = No)
\end{itemize}
To access the data: Copy and paste the dataset from the provided table into a tool of your choice.

\textbf{Dataset:}

\begin{center}
\begin{tabular}{|c|c|c|c|c|c|}
\hline
\textbf{Client ID} & \textbf{Sale Completed} & \textbf{Sales Amount (\$)} & \textbf{No. of Employees} & \textbf{Lead Score} & \textbf{Partner Assigned} \\
\hline
101 & 0 & 0 & 316 & 0.57 & 1 \\
102 & 0 & 0 & 144 & 0.69 & 1 \\
103 & 1 & 10699 & 30 & 0.70 & 1 \\
104 & 1 & 8987 & 338 & 0.50 & 0 \\
105 & 0 & 0 & 176 & 0.93 & 1 \\
106 & 1 & 12875 & 400 & 88. & 0 \\
107 & 0 & 0 & 200 & 0.55 & 1 \\
108 & 1 & 17456 & 35 & 0.95 & 1 \\
109 & 0 & 0 & 90 & 0.62 & 0 \\
110 & 1 & 19342 & 500 & 0.91 & 1 \\
111 & - & 0 & 60 & 0.50 & 0 \\
112 & - & 13452 & 110 & 83. & 1 \\
113 & - & 0 & 45 & 0.65 & 0 \\
114 & - & 14789 & 320 & 0.87 & 1 \\
115 & - & 0 & 75 & 0.59 & 0 \\
116 & 1 & 16004 & 50 & 86. & 1 \\
117 & 0 & 0 & 90 & 0.57 & 0 \\
118 & 1 & 21089 & 280 & 0.90 & 1 \\
119 & 1 & 9875 & 60 & 81. & 1 \\
120 & 0 & 0 & 100 & 0.63 & 0 \\
\hline
\end{tabular}
\end{center}

\textbf{Previous Analysis}

A previous analyst ran the following simple linear regression model:

$Yi = \beta0 + \beta1 X1 + ui$

Where $Yi$ is a binary variable denoting whether the sale was completed or not, and $X1$ is a binary variable denoting whether a sales partner was assigned or not.

They found that $\beta1$ was large (and positive) and statistically significant, concluding that assigning a sales partner strongly increases the probability of making a sale. Senior management is prepared to expand the sales partner program based on this result.

\textbf{Your Tasks}
\begin{itemize}
    \item Critique of Previous Analysis
    \item Identify any possible data quality issues
    \item Explain whether the previous analyst's conclusions were justified based on the methodology used.
    \item Outline an appropriate approach for re-examining the impact of sales partners, specifying the correct statistical model and key variables to include.
\end{itemize}

\textbf{Expected Output}

\textbf{Written Analysis Memo (250-300 words)}
\begin{itemize}
    \item Summarize why the original analysis may have been misleading.
    \item Outline a better regression approach and explain how it might change the conclusions.
    \item Explain the intuition of why this extra review should be conducted to a non-technical management team
\end{itemize}
Note: You do not need to actually perform the regression analysis; only outline the steps needed to conduct a proper analysis. Senior management will use your findings to determine whether to increase, decrease, or refine the use of sales partners in future sales strategies.

\clearpage

\subsubsection*{Consulting Task One: Digital Transformation – AI Adoption in Financial Services}

\textbf{Background}

Evermark Capital, a mid-sized investment management firm, is under pressure to integrate AI-driven investment tools to compete with digital-first investment platforms. While the firm has historically provided high-touch, personalized financial advising, younger clients are shifting to automated investment solutions, and industry trends indicate rapid AI adoption in wealth management.

Leadership wants to introduce AI tools to improve efficiency and expand its customer base, but many senior financial advisors strongly resist AI integration, fearing it will erode trust and reduce their role. The challenge is to adopt AI without alienating existing clients or advisors.

Evermark Capital is considering AI adoption in three areas:
\begin{itemize}
    \item \textbf{Automated Portfolio Management} – AI-driven recommendations based on client risk tolerance and market trends.
    \item \textbf{Client Sentiment Analysis} – AI detects early warning signs of client dissatisfaction by analyzing call transcripts, emails, and portfolio changes.
    \item \textbf{Market Prediction Algorithms} – AI models forecast high-potential investment opportunities based on historical trends and real-time market signals.
\end{itemize}
However, AI adoption comes with operational, regulatory, and reputational risks, and leadership has requested a structured plan for implementation.

\textbf{Key Business \& Market Data}

\textit{Current Business Model:}
\begin{itemize}
    \item Focus on personalized advisory services with an 85\% client retention rate.
    \item Wealthier clients prefer human advisors, while younger clients expect digital solutions.
\end{itemize}

\textit{AI Investment Trends:}
\begin{itemize}
    \item AI-driven platforms are growing 12\% annually, while traditional firms are growing 4\%.
    \item Some competitors offer fully automated investing with fees 30\% lower than Evermark’s advisory services.
\end{itemize}

\textit{Advisor Resistance \& Concerns:}
\begin{itemize}
    \item 58\% of senior advisors worry AI will erode client trust.
    \item Only 27\% of advisors feel comfortable using AI-driven recommendations.
    \item Some advisors believe AI will commoditize financial planning, reducing differentiation.
\end{itemize}

\textit{Regulatory \& Compliance Risks:}
\begin{itemize}
    \item AI-generated financial advice is not yet fully regulated, creating liability risks.
    \item Any errors in AI-driven decisions could lead to compliance violations or client lawsuits.
\end{itemize}

\textit{Technology Costs:}
\begin{itemize}
    \item AI integration requires a \$20M upfront investment and \$5M annually in maintenance.
    \item Expected long-term cost savings: 20\% reduction in advisory labor costs over five years.
\end{itemize}

\textbf{Your Tasks}

\textbf{1. AI Integration Plan \& Rollout Strategy (300-400 words)}

Evermark wants a balanced AI integration plan—one that leverages AI for efficiency and insights without replacing or diminishing the role of human advisors.

Your task is to develop an AI rollout strategy that:
\begin{itemize}
    \item Clearly defines which tasks should be automated and which must remain human-led.
    \item Outlines a phased AI implementation approach, starting with low-risk areas before moving to complex decision-making.
    \item Ensures that AI complements human expertise, rather than replacing advisors, by enhancing their ability to provide tailored investment strategies.
    \item Includes a strategy for client communication, reassuring them that AI enhances decision-making without sacrificing personalized service.
\end{itemize}

\clearpage

\subsubsection*{Consulting Task Three: Crisis Management – Reputation Damage Control}

\textbf{Background}

NaturaCosmetics, a mid-tier skincare brand, is facing a major public relations crisis after a viral social media post alleged that one of its products caused severe skin irritation. The video, posted by a high-profile beauty influencer, has gained 2.3 million views on TikTok and is continuing to spread. As a result, online sales have dropped by 18\% in the past 48 hours, and several customers have expressed concerns across social media platforms.

The company’s leadership has convened an emergency meeting to decide how to respond. While NaturaCosmetics has not had any confirmed safety violations, there have been five similar customer complaints about the PureGlow Night Cream in the past year. These incidents were handled individually but did not previously attract widespread attention.

\textbf{Key Challenges \& Considerations}
\begin{itemize}
    \item \textbf{Consumer Trust Erosion:} Without a well-executed response, this crisis could lead to long-term damage to NaturaCosmetics’ brand reputation.
    \item \textbf{Regulatory \& Legal Risks:} While no official product recalls have been issued, regulatory agencies may begin an investigation if more complaints surface.
    \item \textbf{Retail \& Distribution Concerns:} Large retailers that stock NaturaCosmetics products, such as Sephora and Ulta, are monitoring the situation closely and may reconsider shelf space if the controversy escalates.
    \item \textbf{Competitor Actions:} Rival skincare brands have started subtly marketing their own "dermatologist-approved" products in response to the controversy.
\end{itemize}

\textbf{Your Task}

\textbf{Internal Crisis Strategy Memo (350-500 words):}
\begin{itemize}
    \item \textbf{Immediate Response Plan:} Steps to contain the backlash, including engagement with customers, potential influencer partnerships, and legal considerations.
    \item \textbf{Long-Term Reputation Recovery Strategy:} Proposals for improving product credibility.
    \item \textbf{Risk Assessment of Response Options:} Evaluate different potential actions and their implications.
\end{itemize}
Your response should balance damage control with transparency, ensuring NaturaCosmetics does not escalate the crisis further while rebuilding trust with customers.

\clearpage

\subsubsection*{Manager Task Three: High-Stakes Vendor Selection \& Negotiation}

\textbf{Background}

MetroBuild Infrastructure, a construction project management firm, is preparing to bid on a \$50 million government contract to build a new transit hub. The success of this bid depends heavily on selecting the right steel supplier, as material costs, reliability, and delivery times will directly impact the project timeline and profit margins.

As the Procurement Manager, you are responsible for selecting the best vendor from three shortlisted suppliers. Additionally, your CEO has tasked you with negotiating better payment terms before finalizing the deal.

\textbf{Vendor Options \& Key Considerations}

\begin{center}
\begin{tabular}{|p{2.5cm}|p{2cm}|p{3cm}|p{3cm}|p{3cm}|}
\hline
\textbf{Vendor} & \textbf{Price per Ton (\$)} & \textbf{Delivery Reliability} & \textbf{Contract Terms} & \textbf{Additional Notes} \\
\hline
Apex Metals & 1,500 & High (98\% on-time) & 30\% upfront, balance on delivery & Strong industry reputation, but least flexible on pricing. \\
\hline
Titanium Corp & 1,350 & Medium (90\% on-time) & 50\% upfront, balance over 6 months & Lower price but has had some delays in past projects. \\
\hline
Delta Steel & 1,420 & High (97\% on-time) & 20\% upfront, balance over 12 months & Competitive pricing and favorable payment terms, but a newer player in the industry. \\
\hline
\end{tabular}
\end{center}

\textbf{Additional Information:}
\begin{itemize}
    \item Project delays of more than 2 weeks will result in heavy penalties in the government contract.
    \item The finance team is pushing for lower upfront payments to free up working capital.
    \item The CEO wants a formal negotiation strategy before engaging with suppliers.
\end{itemize}

\textbf{Expected Output}

\textbf{1. Vendor Selection Memo (400-500 words)}
\begin{itemize}
    \item Analyze the pros and cons of each vendor.
    \item Justify your final selection based on cost, reliability, and strategic benefits.
\end{itemize}

\end{document}